\documentclass[iop, apj]{emulateapj}
\usepackage{amsmath}
\usepackage{rotating}

\usepackage{color,hyperref}

\usepackage{color}

\shorttitle{Tidally-driven Roche-Lobe Overflow of Hot Jupiters with MESA}
\shortauthors{Valsecchi, et al.}

\begin{document}

\title{Tidally-driven Roche-Lobe Overflow of Hot Jupiters with MESA}

\author{Francesca Valsecchi\altaffilmark{1}, Saul Rappaport\altaffilmark{2}, Frederic A. Rasio\altaffilmark{1}, Pablo Marchant\altaffilmark{3}, \and Leslie A. Rogers\altaffilmark{4}}

\altaffiltext{1}{Center for Interdisciplinary Exploration and Research in Astrophysics (CIERA), and Northwestern University, Department of Physics and Astronomy, Evanston, IL 60208, USA. FV: francesca@u.northwestern.edu; FR: rasio@northwestern.edu.}
\altaffiltext{2}{Department of Physics, and Kavli Institute for
  Astrophysics and Space Research,  Massachusetts Institute of
  Technology, Cambridge, MA 02139, USA, sar@mit.edu.}
\altaffiltext{3}{Argelander-Institut fŸr Astronomie, UniversitŠt Bonn, Auf dem HŸgel 71, D-53121 Bonn, Germany; pablo@astro.uni-bonn.de.}
\altaffiltext{4}{Department of Astronomy and Department of Geophysics and Planetary Sciences, 
California Institute of Technology, Pasadena, CA 91125, USA; larogers@caltech.edu.}

%\author{F. Valsecchi, F. A.\ Rasio}
%\affil{Center for Interdisciplinary Exploration and Research in
%  Astrophysics (CIERA)} 
%\affil{Department of Physics and Astronomy,
%  Northwestern University, Evanston, IL 60208, USA} 
%
%
%\author{S. \ Rappaport}
%\affil{Department of Physics, and Kavli Institute for
%  Astrophysics and Space Research, \\
%  Massachusetts Institute of
%  Technology, Cambridge, MA 02139, USA, sar@mit.edu}
%
%
%\author{P. \ Marchant}
%\affil{Instituto de Astrof'sica, Facultad de F'sica, Pontificia Universidad Cat—lica de Chile, \\
%Av. Vicu–a Mackenna 4860, Macul 7820436, Santiago, Chile; \\
%Argelander-Institut fŸr Astronomie, UniversitŠt Bonn \\
%Auf dem HŸgel 71, D-53121 Bonn, Germany; pablo@astro.uni-bonn.de}
%
%\and
%
%\author{L. \ Rogers}
%\affil{Department of Astronomy and Department of Geophysics and Planetary Sciences, \\
%California Institute of Technology, Pasadena, CA 91125, USA; larogers@caltech.edu}

\begin{abstract}
Many exoplanets have now been detected in orbits with ultra-short periods, very close to the Roche limit. Building upon our previous work, we study the possibility that mass loss through Roche lobe overflow (RLO) may affect the evolution of these planets, and could possibly transform a hot Jupiter into a lower-mass planet (hot Neptune or super-Earth). We focus here on systems in which the mass loss occurs slowly (``stable mass transfer'' in the language of binary star evolution) and we compute their evolution in detail with the binary evolution code MESA. We include the effects of tides, RLO, irradiation and photo-evaporation of the planet, as well as the stellar wind and magnetic braking. Our calculations all start with a hot Jupiter close to its Roche limit, in orbit around a sun-like star. The initial orbital decay and onset of RLO are driven by tidal dissipation in the star. We confirm that such a system can indeed evolve to produce lower-mass planets in orbits of a few days. The RLO phase eventually ends and, depending on the details of the mass transfer and on the planetary core mass, the orbital period can remain around a few days for several Gyr. The remnant planets have a rocky core and some amount of envelope material, which is slowly removed via photo-evaporation at nearly constant orbital period; these have properties resembling many of the observed super-Earths and sub-Neptunes. For these remnant planets we also predict an anti-correlation between mass and orbital period; very low-mass planets ($M_{\rm pl}\,\lesssim\,5\,M_{\oplus}$) in ultra-short periods ($P_{\rm orb}\,\textless\,1\,$d) cannot be produced through this type of evolution.
\end{abstract}

\section{Introduction} \label{Intro}
Hot Jupiters, giant planets in orbits of a few days, constitute one of the many surprises of exoplanet searches. Whether their tight orbits are the result of inward migration in a protoplanetary disk \citep{GoldreichTremaine80,Lin+96,Ward97,MurrayHHT98}, or tidal circularization of an orbit made highly eccentric via gravitational interactions \citep{RasioFord96,WuMurray03,FabryckyTremaine07, ChatterjeeFMR08,Nagasawa08,WuLithwick11,Naoz+11,PlavchanBilinski13, ValsecchiR2014b, ValsecchiR2014}, is still matter of debate.  Certainly, independent of the formation mechanism, tidal dissipation in the slowly-spinning host stars is causing the orbits of the tightest hot Jupiters currently known to shrink rapidly (e.g., \citealt{RTLL1996,Sasselov2003,Birkby+14, ValsecchiR2014b}; see also Table ~1 in \citealt{ValsecchiR2014b} and references therein). 

Eventually, hot Jupiters may decay down to their Roche-limit separation. While it is commonly assumed that the planet is then quickly accreted by the star (e.g., \citealt{Jackson+09,Metzger+12,SchlaufmanWinn13, DamianiLanza14,TeitlerKonigl14, ZhangPenev14}), for a typical system hosting a hot Jupiter orbiting a sun-like star the ensuing mass transfer (hereafter `MT') may be dynamically stable \citep{SepinskyWKR10}. This was suggested, e.g., to explain WASP-12's transit features \citep{LaiHvdH2010}. \cite{TrillingBGLHB1998} investigated the possibility of halting inward disk migration through tides and Roche-lobe overflow (`RLO') MT from a hot Jupiter to a rapidly-spinning (young) stellar host. However, we note that the host stars of the tightest hot Jupiters are observed to be rotating slowly at present.

In \cite{ValsecchiRS2014}, we investigated the fate of  a hot Jupiter transferring mass to its stellar host using a simplified binary MT model. We showed that the planet could be stripped of its envelope, resulting in a hot super-Earth-type planet. This model naturally solves some of \emph{Kepler}'s evolutionary puzzles (e.g., Kepler-78; \citealt{Howard+13Nat, Pepe+13Nat,Sanchis-OjedaRWLKLB2013}), and it could explain the excess of isolated hot super-Earth- or sub-Neptune-size planets seen in the \emph{Kepler} data \citep{Steffen:2013c}. 

Our previous work relied on several simplifying and potentially key assumptions. 
First, while using detailed models for the host star, we adopted published mass-radius relations for the planetary component, thus assuming that the planet remains in thermal equilibrium throughout the MT. Second, even though our planetary models included the effect of stellar irradiation on the  planetary mass-radius relations, irradiation was kept \emph{fixed}, while it is expected to vary as the orbit evolves during MT. Furthermore, we neglected the resulting mass loss due to photo-evaporation \citep{Murray-Clay+09,JacksonMBRFG2010,LopezFortney12,LopezFortney13,OwenAlvarez2015}, even though various studies found it to play an important role in the evolution of highly irradiated super-Earth and sub-Neptune-type planets (e.g., \citealt{JacksonMBRFG2010,RogersBLS2011, LopezFortney12,BatyginStevenson13,LopezFortney13,OwenWu2013}). 

As a natural continuation of our previous work, here we significantly expand upon our simple binary MT model to include detailed planetary evolution, as well as the effects of a varying irradiation and the consequent photo-evaporative mass loss from the planet.  For these new calculations we utilize the Modules for Experiments in Stellar Astrophysics (MESA) evolution code \citep{PaxtonBDHLT2011,PaxtonCABBDMMSTT+13,PaxtonMSBBCDFHLTTT2015}. The MESA inlist files used in this work can be downloaded at 
\href{https://github.com/FrancescaV/planet\_MT\_with\_MESA\_inputs\_and\_models}{https://github.com/FrancescaV/planet\_MT\_with\_MESA}
\\
\_inputs\_and\_models.

The plan of the paper is as follows. In Section \S~\ref{Input Physics} we describe the stellar and planetary models used in this work. In Section \S~\ref{Orbital Evolution Model} we describe our orbital evolution model and the physical mechanisms entering the calculation. We present some examples of our orbital evolution calculations in Section~\ref{Example} and discuss our results in Section~\ref{discussion}. We conclude in Section~\ref{Conclusions}.

For quick reference, the notations adopted in this work are summarized in Table~\ref{Tab:ParamsDefinition}.
%%%%%%%%%%%%%%%%%%%%%%%%%%
\begin{table}[!h]
\centering
\caption{Definition of the main parameters used in this work.}
\begin{tabular}{lr}
\hline 
{\bf Parameter} & {\bf Definition} \\
\\[-1.0em]
\hline \\[-1.0em]
\smallskip
$M_{*}$, $M_{\rm pl}$ & Mass\\
\smallskip
$M_{\rm c}$, $M_{\rm env}$ & Planetary core and envelope mass\\
\smallskip
$R_{*}$, $R_{\rm pl}$ & Radius\\
\smallskip
$R_{\rm lobe,*}$, $R_{\rm lobe, pl}$ & Roche-lobe radius\\
\smallskip
$T_{\rm *}$, $T_{\rm pl}$ & Surface temperature\\
\smallskip
$\Omega_{*}$, $\Omega_{\rm pl}$ ($\Omega_{\rm o}$) & Spin (orbital) frequency\\
\smallskip
$Z$ & Metallicity\\
\smallskip
$a$ & Semimajor axis\\
\smallskip
$P_{\rm orb}$ & Orbital period\\
\smallskip
$J_{\rm orb}$ & Orbital angular momentum\\
\smallskip
$f\,=\,M_{\rm env}/M_{\rm pl}$ & Envelope mass fraction\\
\smallskip
$q\,=\,M_{\rm pl}/M_{\rm *}$ & Planet to star mass ratio\\
\smallskip
$t$, $t_{\rm MS}$ & Stellar age and main-sequence lifetime\\
\\[-1.0em]
\hline \\[-1.0em]
\end{tabular}
\label{Tab:ParamsDefinition}
\tablecomments{The subscripts ``*'' and ``pl'' refer to the star and planet, respectively. We take the stellar age to be equal to the system age, and the main-sequence lifetime for solar mass stars with radiative cores to be the age when the mass fraction of $H$ at the center of the star $\rightarrow 0$.}
\end{table}
%%%%%%%%%%%%%%%%%%%%%%%%%%

%%%%%%%%%%%%%%%%%%%%%%%%%%%%%%%%%%%%%%%%%%%%%%%%
\section{Stellar and Planetary Models}\label{Input Physics}

The stellar and planetary models adopted in this work are computed with MESA (version 7184; ~\citealt{PaxtonBDHLT2011,PaxtonCABBDMMSTT+13,PaxtonMSBBCDFHLTTT2015}). In particular, the planets are created and evolved closely following the test suite  \emph{make\_planets} provided within MESA and the input files yielding Figure~3 of ~\cite{PaxtonCABBDMMSTT+13}\footnote{available at $http://mesastar.org/results/mesa2/planets$}. 
In what follows we give specifications only for those MESA parameters that are changed from the values adopted in these input files.

In all our calculations, we consider a 1$\,M_{\odot}$ star paired with a 1$\,M_{\rm J}$ planet. We assume solar composition ($Y\,=\,0.27$, $Z\,=\,0.02$) for both the star and planet envelope, and keep the mixing length parameter $\alpha_{\rm MLT}$ to MESA's default value of 2. 
For the planet, we expand on our previous work \citep{ValsecchiRS2014} and consider models with solid cores of masses $M_{\rm c}\,=\,1\,M_{\oplus}, 5\,M_{\oplus}, 10\,M_{\oplus}, 15\,M_{\oplus}$ and, 30$\,M_{\oplus}$. 
For the cores, we use a constant density of 5\,g\,cm$^{-3}$, following \cite{BatyginStevenson13}. Furthermore, we take the heat-flux arising from radioactive decay in the core to be zero, following \cite{FortneyMB07}.
As these authors point out, this is a common assumption in evolutionary models of Jupiter- and Saturn-type planets (\citealt{Hubbard1977,SaumonEtAl1992,FortneyH2003}), as this approach introduces a small error compared to other unknowns entering the problem. However, we note that \cite{LopezFortney2014} found such heating to play an important role in delaying cooling and contraction, particularly for planets less than 5$M_{\oplus}$. This could lead to an underestimate of the radii of sub-Neptune planets, especially at ages $\leq\,$1\,Gyr. Below we focus on planetary models at least 2\,Gyr old.

\begin{figure} [!h]
\epsscale{1.0}
\plotone{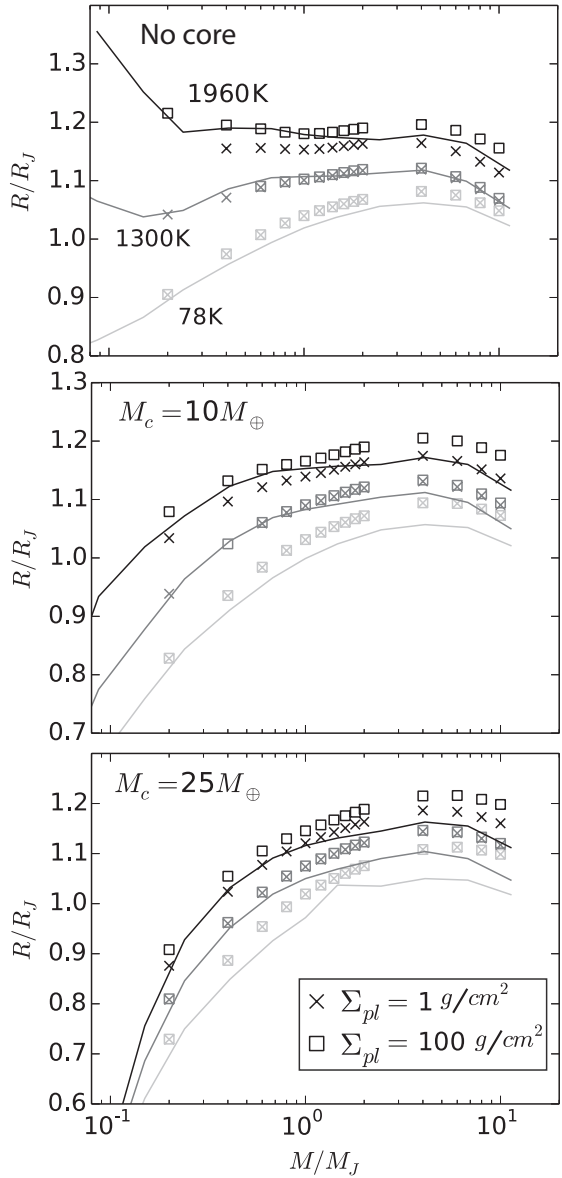}
\caption{Planetary mass-radius relations at 4.5\,Gyr for different core masses and equilibrium temperatures. From top to bottom: core-less planets, $M_{\rm c}\,=\,10\,M_{\oplus}$, and 25$\,M_{\oplus}$, as illustrative examples. The solid lines are the models by \cite{FortneyMB07} (from Table~4 of their paper) at an equilibrium temperature of 78\,K (light grey), 1300\,K (dark grey), and 1960\,K (black). The data points are the models computed with MESA for $\Sigma_{\rm pl}$ set to 1\,g/cm$^{2}$ ($``\times''$) and 100\,g/cm$^{2}$ (``$\square\,$''). Varying $\Sigma_{\rm pl}$ between 0.1$-$1\,g/cm$^{2}$ makes no significant difference. The value of $\Sigma_{\rm pl}$ is not important for $T_{\rm eq}\leq1300K$ (symbols overlap). As in \cite{FortneyMB07}, the radii correspond to a pressure of $\sim$1\,bar. We note that the initial value of %$radius\_in\_cm\_for\_create\_initial\_model$ 
the planetary radius required by the $create\_initial\_model$ routine (see Section~2.1 of \citealt{PaxtonCABBDMMSTT+13}) was varied between 2$-$5\,$R_{\rm J}$ to facilitate MESA's convergence.}
\label{fig:mass_radius_compareFortney_irradiation_radiusAt1bar} 
\end{figure}

We account for the effects of irradiation and photo-evaporation as follows.
For irradiation, we use the $F_{*}-\Sigma_{\rm pl}$ surface heating function. Here $F_{*}$ is the day-side flux from the star at the substellar point, given by
\begin{eqnarray}
F_{*} = \sigma \,T_{*}^{4} \left(\frac{R_*}{a}\right)^2. % 4\,\sigma\,T_{\rm eq}^{4}, 
\label{eq:Fstar}
\end{eqnarray}
The planet equilibrium temperature, $T_{\rm eq}$, is 
\begin{eqnarray}
T_{\rm eq}\,=\,T_{*}\left(\frac{R_{*}}{2a}\right)^{1/2},
\label{eq:Teq}
\end{eqnarray}
\citep{SaumonHBGLC1996} so that the power received from the host star could be radiated away in equilibrium if the planet had this  temperature over its entire surface.  The parameter $\Sigma_{\rm pl}$ is the column depth reached by irradiation. Here we adopt a value of $\Sigma_{\rm pl}$\,=\,1\,g/cm$^{2}$, which yields planetary mass-radius relations in agreement with detailed models by~\cite{FortneyMB07}, within a few percent for highly irradiated planets, as shown in Figure~\ref{fig:mass_radius_compareFortney_irradiation_radiusAt1bar}. We discuss the effect of varying $\Sigma_{\rm pl}$ in Section~\ref{varing sigma}. In what follows, the planetary radius corresponds to an optical depth $\tau\,=\,2/3$.

%%%%%%%%%%%%%%%%%%%%%%%%%%%%%%%%%%%%%%%%%%%%%%%%
%%%%%%%%%%%%%%%%%%%%%%%%%%%%%%%%%%%%%%%%%%%%%%%%
Irradiation leads to photo-evaporative (`PE') mass loss from the planet. Specifically, this process is thought to be most efficient when a hot Jupiter is strongly irradiated by ultraviolet (UV) and X-ray photons, which photo-ionize atomic $H$ in the planetary atmosphere. The resulting heat input, when high enough to induce temperatures corresponding to the escape velocity, can cause outflows. \cite{Murray-Clay+09} identified two regimes, based on the stellar flux $F_{\rm XUV}$ (however, see \citealt{OwenAlvarez2015}). 
For large $F_{\rm XUV}$, like those typical of T-Tauri stars ($\,\gtrsim\,10^{5}$\,erg\,cm$^{-2}$\,s$^{-1}$)
% \,>\,10^{4}$\,erg\,cm$^{-2}$\,s$^{-1}$ 
the mass loss is ``radiation/recombination'' limited and it is described by
\begin{align}
\dot{M}_{\rm rr-lim}\sim\,4\times\,10^{12}\left(\frac{F_{\rm XUV}}{5\times\,10^{5}\,{\rm erg\,cm}^{-2}\,{\rm s}^{-1}}\right)^{1/2}\,{\rm g\,s}^{-1}.
\label{eq:MdotPE_rrLim}
\end{align}
For lower $F_{\rm XUV}$ values the mass loss is ``energy limited'' and it is described by \citep{ErkaevKLSLJB07,LopezFortney12}
\begin{align}
\dot{M}_{\rm e-lim}\approx\,\frac{\epsilon\pi F_{\rm XUV}R_{\rm XUV}^{3}}{GM_{\rm p}\,K_{\rm tide}},
\label{eq:MdotPE_eLim}
\end{align}
where $K_{\rm tide}=1-(3/2)(1/\xi)+(1/2)(1/\xi^{3})$ and $\xi=R_{\rm Hill}/R_{\rm XUV}$.
Since in all our calculations $F_{\rm XUV}$ (computed as described below) remains below $10^{5}{\rm erg\,cm}^{-2}\,{\rm s}^{-1}$, we use only Eqn.~(\ref{eq:MdotPE_eLim}). However, we discuss whether our results are sensitive to the photo-evaporation prescription in Section~\ref{Varying Photo-Evaporation} and Table~\ref{Tab:SummaryResults}. 

For the flux, we follow \cite{RibasGGA05} and take 
\begin{align}
F_{\rm XUV}\,=\,29.7 \left(\frac{t}{\rm Gyr}\right)^{-1.23}\, \left(\frac{a}{{\rm AU}}\right)^{-2}\,{\rm erg\,s}^{-1}\,{\rm cm}^{-2},
\label{eq:Fxuv}
\end{align}
where we have scaled their result at 1 AU to an arbitrary distance. %Eqn.~(\ref{eq:Fxuv}) is valid at 1AU. 
The parameter $\epsilon$ represents the efficiency of converting $F_{\rm XUV}$ into usable work, while $R_{\rm XUV}$ is the radius of the planet at which the atmosphere becomes optically thick to XUV photons.  \cite{Murray-Clay+09} place $R_{\rm XUV}$ at a surface pressure of $\sim\,10^{{-9}}\,$bar, which corresponds to a radius that is typically 10\%$-$20\% greater than the optical photosphere~(\citealt{LopezFortney13}, hereafter LF13). 
Here we closely follow the recent results of LF13 and adopt $\epsilon\,=\,0.1$. We note, however, that hydrodynamic calculations suggest that $\epsilon$ can vary between 0.01$-$0.2, depending on the mass of the planet~\citep{OwenJackson2012}. With $\epsilon\,=\,0.1$, the $R_{\rm XUV}$ value that better matches the results of detailed numerical calculations by LF13 for systems 1$-$10\,Gyr old is $R_{\rm XUV}\,=1.2\,R_{\rm pl}$ (see below) and we use this estimate of $R_{\rm XUV}$ throughout our calculations\footnote{MESA uses automatic mesh refinement, thus adjusting the number of mesh points of the planetary model at the beginning of each timestep, if necessary. For this reason, the point at the surface where the pressure is $\sim\,10^{{-9}}\,$bar is not always resolved.}. 
Finally, $K_{\rm tide}$ is to account for the fact that the mass leaving the planet needs only to reach the Hill radius to escape, where $R_{\rm Hill} = M_{\rm pl}^{1/3}(3M_{\rm *})^{-1/3}a$~\citep{ErkaevKLSLJB07}. 

\begin{figure} [!h]
\epsscale{1.1}
\plotone{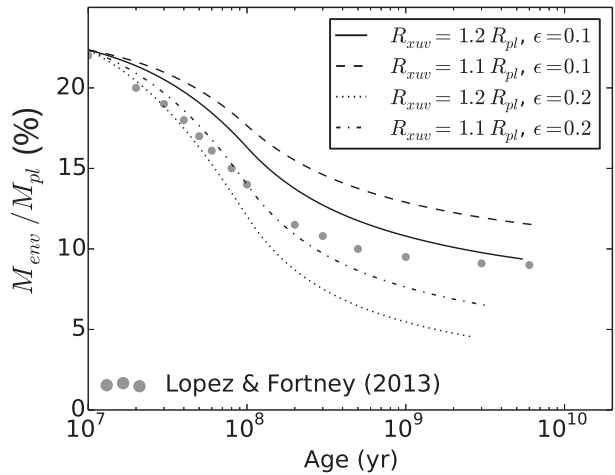}
\caption{Mass loss evolution of Kepler-36c. Percent of mass contained in the envelope as a function of time for different $R_{\rm XUV}$ and $\epsilon$ values. The grey data points are taken from Figure~1 of LF13. The parameter $\Sigma_{\rm pl}$ was set to 1\,g\,cm$^{-2}$. For $F_{\rm XUV}$, we follow \cite{RibasGGA05}, but rescale the flux at the Kepler-36c orbital separation ($a=0.12\,$AU). As in LF13, we keep $F_{\rm XUV}$ constant at the 100\,Myr value when the star is younger than 100\,Myr and we let it evolve when the star is older than 100\,Myr. Here we assume a 1\,$M_{\odot}$ companion. At the surface ($\tau\,=\,2/3$), the pressure is $\sim\,10\,$mbar (where LF13 place the transiting radius).}
\label{fig:Fig1_LopezFortney2013_sigma1} 
\end{figure}

We test the implementation of photo-evaporation by comparing with the mass loss calculations presented by LF13 for the planet Kepler-36c. These are shown in Figure~\ref{fig:Fig1_LopezFortney2013_sigma1}.
Following their work, we create a 9.4\,$M_{\oplus}$, 10\,Myr old irradiated planet  with a core of 7.4\,$M_{\oplus}$ and Z\,=\,0.35. We adopt $T_{\rm eq}=930\,$K (\citealt{CarterEtAl2012} reports $T_{\rm eq}\,=\,928\,\pm\,10\,$K) and evolve the planetary model with irradiation (fixed) and photo-evaporation, according to Eqn.~(\ref{eq:MdotPE_eLim}). The parameters adopted in this work are those yielding agreement with LF13 within a few percent for 1$-$10\,Gyr old systems (black solid line).
%%%%%%%%%%%%%%%%%%%%%%%%%%%%%%%%%%%%%%%%%%%%%%%%
%%%%%%%%%%%%%%%%%%%%%%%%%%%%%%%%%%%%%%%%%%%%%%%%%%%%%%%%%%%%%%%%%%%%%%%%%%%
\section{Orbital Evolution Model}\label{Orbital Evolution Model}
MESA allows us to track the evolution of both the planet and star simultaneously, while orbital evolution is followed by taking into account changes to the orbital angular momentum of the system. Below we describe the physical mechanisms included in our model: photo-evaporation (`PE'), tides,
% \emph{in the star}
 magnetic braking (`MB'; \citealt{Skumanich72}), and RLO.  
For simplicity, we neglect the effects of stellar wind mass loss. From the wind prescription provided in MESA's test suite \emph{1M\_pre\_ms\_to\_wd} for the evolution of a 1$M_{\odot}$ star (\citealt{Bloecker95,Reimers75}), we find that the orbital evolution timescales associated with stellar winds are longer than $10^{12}$\,yr throughout the main-sequence evolution of the host star.

 %{\bon We also present results both with and without stellar wind mass loss, as a comparison.}
% to \cite{ValsecchiRS2014}, we neglect the effects of stellar wind mass loss on the orbit and spin of the star, as we expect it not to be significant during the main sequence lifetime of the stellar component, but we discuss it in Section~\ref{discussion}.}
%%%%%%%%%%%%%%%%%%%%%%%%%%%%%%%%%%%%%%%%%%%%%%%%
\subsection{Photo-Evaporation}\label{Photo-Evaporation}
Mass escape via photo-evaporation affects the orbital separation and the spin of the planet. 
%We do not account for its effects on the negligible planetary spin. 
For simplicity, we assume spherically symmetric mass loss, which carries away the specific angular momentum of the mass losing component. With this assumption, photo-evaporation affects the orbital angular momentum $J_{\rm orb}$ according to
\begin{align}
\left(\frac{\dot{J}_{\rm orb}}{J_{\rm orb}}\right)_{PE}&=\frac{\dot{M}_{\rm pl,PE}}{M_{\rm pl}}\frac{1}{1+q} \simeq\frac{\dot{M}_{\rm pl,PE}}{M_{\rm pl}}, 
\label{eq:JdotPE}
\end{align}
where $q\,=\,M_{\rm pl}/M_{*}$. For reference, the evolution of the orbital separation due to photo-evaporation is given by
\begin{align}
\left(\frac{\dot{a}}{a}\right)_{PE}&=-\frac{\dot{M}_{\rm pl,PE}}{M_{\rm pl}}\frac{q}{1+q} \simeq\,0, 
\label{eq:a_dotPE}
\end{align}

Photo-evaporation is always active and $\dot{M}_{\rm pl,PE}$ is given by either Eqn.~(\ref{eq:MdotPE_rrLim}) or Eqn.~(\ref{eq:MdotPE_eLim}), depending on $F_{\rm XUV}$ though, in practice, only Eqn.~\ref{eq:MdotPE_eLim} is needed in our calculations.
%\footnote{\ron Where in the paper do we say that, except for special tests, we used only Eqn.~(4)?}.
As far as the planetary spin is concerned, photo-evaporation carries away the angular momentum of the corresponding shells of material.

While we assume that mass loss is spherically symmetric, we note that strong magnetic fields may confine the flow primarily to the poles and day-side of the planet \citep{OwenAdams2014}. \cite{TeyssandierEtAl2015} studied the torque on super-Earth and sub-Neptune-type planets due to anisotropic photo-evaporative mass loss using steady-state  one-dimensional wind models. They found that only in rare cases is the planet's orbit affected by wind torques. 
%%%%%%%%%%%%%%%%%%%%%%%%%%%%%%%%%%%%%%%%%%%%%%%%
\subsection{Tides}
Tides affect the stellar and planetary spins, as well as the orbital separation, transferring angular momentum between the orbit and the components' spin. 
%striving to drive the components to synchronism by exchanging angular momentum between their spins and the orbit. Therefore, tides shrink (expand) the orbit whenever the spin frequency is smaller (bigger) than the orbital frequency. 
Within MESA, tides are first applied to the spins. The orbital separation is then varied so as to conserve total angular momentum.
Here we take the components to be rotating as solid bodies (however, see \citealt{Stevenson1979,BarkerDL2014}). 

%$\Omega_{*}/\Omega_{\rm o}<1$ ($\Omega_{*}/\Omega_{\rm o}>1$). 
For \emph{stellar tides} we proceed as in \cite{ValsecchiRS2014} and \cite{ValsecchiR2014b,ValsecchiR2014}. Specifically, we adopt the weak-friction approximation \citep{Zahn1977, Zahn1989} using a parametrization for tidal dissipation calibrated from observations of stellar binaries, as in~\cite{HurleyTP02}.
% (as opposed to using the highly uncertain tidal-$Q$ factor\,\citealt{GoldreichSoter1966})
This assumes that tides are dissipated in the stellar convection zone via eddy viscosity. Furthermore, we reduce the efficiency of tides at high tidal forcing frequencies (when the forcing frequency is higher than the convective turnover frequency of the largest eddies) linearly, following \cite{Zahn1966}, and as suggested by recent numerical results by \cite{PenevSRD2007}. 

For \emph{planetary tides}, we assume they efficiently maintain the planet in a tidally locked configuration ($\Omega_{\rm pl}=\Omega_{\rm o}$) throughout the evolution. In fact, for a tidal quality factor $Q'\,=\,10^{6}$ (typical for gas giants) and a nearly synchronized planet\footnote{To get a sense for the magnitude of the spin synchronization timescale we use Eqn.~(10) in \cite{MatsumuraPR2010} and $\Omega_{\rm pl}(t)=\Omega_{\rm o}(t-\Delta t)$, where $\Delta\,t$ is the time interval between two consecutive time steps during an orbital evolution calculation.}, the spin synchronization timescale due to static tides in the planet is $\sim\,2-4$ orders of magnitude shorter than the timescale related to the main driver of the orbital evolution (i.e., mass loss from the planet; see Section~\ref{Example}), depending on the core mass. Clearly, tides would synchronize the planet even faster for lower values of $Q'$ more appropriate for rocky planets. We  further discuss tidal locking for the planet in Section~\ref{discussion}.

Even though our calculations account for tides in both components, our results show that only stellar tides can affect the orbital separation significantly. In particular, for a slowly spinning stellar host, tides transfer angular momentum from the orbit to the stellar spin, causing orbital decay.

%%%%%%%%%%%%%%%%%%%%%%%%%%%%%%%%%%%%%%%%%%%%%%%%
\subsection{Magnetic Braking}
For the loss of stellar spin angular momentum via magnetic braking we follow \cite{Skumanich72} and adopt 
\begin{align}
(\dot{\Omega}_{*})_{MB} = -\alpha_{MB}\Omega_{*}^{3},
\end{align} 
where $\alpha_{MB}=1.5\times 10^{-14}~$yr (e.g., \citealt{BarkerOgilvie2009,DobbsDixon+04,MatsumuraPR2010,ValsecchiR2014b,ValsecchiR2014}). This law is well established for the stellar equatorial rotation rates of interest here (1-30\,km\,s$^{-1}$).
%%%%%%%%%%%%%%%%%%%%%%%%%%%%%%%%%%%%%%%%%%%%%%%%
\subsection{Roche-Lobe Overflow}\label{RLO}
RLO is modeled by implicitly computing the mass transfer rate that is required for the planetary radius to remain below its Roche lobe radius, for which we use the approximation by \cite{Eggleton1983}. This procedure is described in Section~2.3.2 of \cite{PaxtonMSBBCDFHLTTT2015}. Here we consider both conservative and non-conservative MT. 
When MT is conservative, all mass leaving the planet via RLO is accreted onto the star and $\dot{M}_{\rm *}\,=\,-\dot{M}_{\rm pl,RLO}$. Instead, during non-conservative MT evolution, some fraction $\delta$ of $\dot{M}_{\rm pl,RLO}$ is lost from the system and $\dot{M}_{\rm *}\,=\,-(1-\delta)\dot{M}_{\rm pl,RLO}$. Note that in none of the examples presented here does the {\em star} fill its Roche-lobe during the orbital evolution calculation.%Below we explain how the orbit is evolved in the two different MT scenarios.

\subsubsection{Stable Conservative Mass Transfer}\label{Stable Conservative}
During conservative MT,
%($\delta\,=\,0$ and $\dot{M}_{\rm *}\,=\,-\dot{M}_{\rm pl,RLO}$), 
the evolution of $J_{\rm orb}$ is generally computed under the assumption that MT proceeds through an accretion disk. In the standard picture, the matter flowing through the inner Langrangian point appears to be pushed into orbit about the host star by Coriolis forces (as viewed in the corotating frame). Subsequently, viscous stresses spread this material into a disk around the accretor~\citep{FrankKingRaine1985}. This disk transports mass toward the accretor and angular momentum away from it. The latter is eventually returned to the orbit via torques operating between the donor and the outer edge of the disk~\citep{LinPapaloizou1979} and a small residual angular momentum is transferred to the spin of the star.
Our calculations with MESA include spin-up through accretion. However, as the material accreted by the star is only a tiny fraction of its total mass, spin-up due to accretion is negligible.
%Formally, MESA requires information about the specific angular momentum of the accreted material to modify the accretor's spin accordingly. In our calculations we assume that the accreted material has the same angular momentum as a parcel of mass of the surface of the star, i.e., it is approximately co-rotating. As the amount of mass accreted by the star is much smaller than the stellar mass, spin up due to accretion is negligible.

\subsubsection{Stable Non-Conservative Mass Transfer}\label{Stable non-Conservative}
During non-conservative MT, changes in $J_{\rm orb}$ are computed as in \cite{SobermanPV97} 
\begin{align}
\left(\frac{\dot{J}_{\rm orb}}{J_{\rm orb}}\right)_{RLO}&=\delta\gamma(1+q)^{1/2}\frac{\dot{M}_{\rm pl, RLO}}{M_{\rm pl}}\simeq\delta\gamma\frac{\dot{M}_{\rm pl, RLO}}{M_{\rm pl}}. 
\label{eq:JdotRLO}
\end{align}
For comparison, the evolution of the orbital separation due to RLO is given by
\begin{align}
\left(\frac{\dot{a}}{a}\right)_{\rm RLO} \sim \, 2 \frac{\dot{M}_{\rm pl,RLO}}{M_{\rm pl}}(\delta\gamma-1) \, > \,0.
\label{eq:aRLO}
\end{align}

This model assumes that a fraction $\delta$ of $\dot{M}_{\rm pl,RLO}$ settles into a ring whose radius $a_{r}$ is a constant multiple $\gamma^{2}$ of the orbital separation $a$. This mass is then lost from the system, taking with it the specific angular momentum of the ring.  The remaining fraction of the mass $(1-\delta)$ is assumed to be accreted onto the host star.  Below we describe our choices for $\gamma$ and $\delta$, based on the expected stability of MT. 
%This model assumes that $\dot{M}_{\rm pl,RLO}$ settles into a ring whose radius $a_{r}$ is a constant multiple $\gamma^{2}$ of the orbital separation $a$. A fraction $\delta$ of $\dot{M}_{\rm pl,RLO}$  can then be lost from the system. Below we describe our choices for $\gamma$ and $\delta$, based on the expected stability of MT. %For the examples presented in Section~\ref{Example} we set $\delta\,=\,1$ and vary $\gamma$.}
%%%%%%%%%%%%%%%%%%%%%%%%%%%%%%%%%%%%%%%%%%%%%%%%
\subsubsection{Considerations on the Stability of Mass Transfer}
The calculations presented in Section~\ref{Example} show that conservative MT is always stable. Instead, some care must be taken when considering non-conservative MT.

For the systems under consideration, the stability of the MT phase depends mainly on the fraction of planetary mass leaving the system, its specific angular momentum, and the response of the planet to mass loss. For a mass-losing planet the latter cannot be determined a priori and it is computed with MESA as the orbital evolution proceeds. We also have no prior knowledge of the values of $\gamma$ and $\delta$ (the parameters regulating the amount of mass leaving the system and how much specific angular momentum is carried away) that correspond to dynamical stability of MT. However, some guidance on the region of stability can be gained as follows. For systems with extreme mass ratios ($q\ll 1$, such as those considered here) and $\dot{J}_{\rm orb}/J_{\rm orb}$ given by Eqns.~(\ref{eq:JdotPE}) and (\ref{eq:JdotRLO}), the planetary mass during RLO changes according to (e.g., \citealt{Rappaport+82} for a derivation)
\begin{align}
\frac{\left |\dot{M}_{\rm pl, RLO}\right |}{M_{\rm pl}}\simeq\frac{-\left(\frac{\dot{J}_{\rm orb}}{J_{\rm orb}}\right)_{\rm tides}+\frac{1}{2}\left(\frac{\dot R_{\rm pl}}{R_{\rm pl}}\right)_{\rm therm}-\frac{\dot{M}_{\rm pl, PE}}{M_{\rm pl}}(\frac{ 1}{6}-\frac{\xi}{2})}{\frac{\xi}{2} -\delta\gamma+\frac{5}{6}}.
\label{eq:MdotPlanet}
\end{align}
Here $(\dot R_{\rm pl}/R_{\rm pl})_{\rm therm}$ is the fractional rate of change of the planet radius due to its thermal evolution and $(\dot{J}_{\rm orb}/J_{\rm orb})_{\rm tides}$ is the fractional rate of change in orbital angular momentum due to tides. Finally, $\xi\,=\,({\rm d}\,{\ln} R_{\rm pl} /{\rm d}\,{\ln} M_{\rm pl} )_{\rm ad}$ is the planet's adiabatic logarithmic derivative of radius with respect to mass, not to be confused with the mass-radius relations in Figure~\ref{fig:mass_radius_compareFortney_irradiation_radiusAt1bar}, valid for thermal equilibrium models. A necessary condition for stable MT is that the denominator of Eqn.~(\ref{eq:MdotPlanet}) be positive.
 
Our goal here is to improve on the \emph{stable} MT calculations presented in \cite{ValsecchiRS2014} by investigating the importance of irradiation effects and self-consistent models for the planet. To explore MT stability in the present work, we set $\delta\,=\,1$ and consider $\gamma$ values for which the MT is expected to be dynamically stable. There is no loss of generality in this prescription since $\gamma$ and $\delta$ appear only as a product in Eqn.~(\ref{eq:MdotPlanet}). We discuss possible realistic mass transfer configurations  in Section~\ref{discussion}, while reserving a more detailed analysis of MT stability in hot-Jupiter systems to a future study. Such analysis should account for a broad range of $\gamma$ and $\delta$ values, as well as initial orbital configurations and properties of the components (e.g., $M_{*}$, $M_{\rm pl}$, and metallicity; see Section~\ref{Conclusions}). 

With $\delta\,=\,1$ and $\gamma\,\neq\,0$  we are assuming that all RLO material leaves the system carrying away a specific angular momentum equal to $\gamma\sqrt{GM_{*}a}$. 
The condition for stability, i.e., the requirement of a positive denominator in Eqn.~(\ref{eq:MdotPlanet}), then reduces to 
\begin{align}
\gamma<\frac{5}{6}+\frac{\xi}{2}.
\label{eq:MTstability}
\end{align}
Eqn.~(\ref{eq:MTstability}) shows that an increase in the planetary radius with adiabatic mass loss (i.e., $\xi < 0)$ has a destabilizing effect, as expected. 
%Detailed planetary models show that such an increase occurs for highly irradiated planets, and that it is more severe for cores of few Earth mass (e.g., ~\citealt{FortneyMB07,BatyginStevenson13}). 
%Instead, for an $n\,=\,1$ polytropic profile for the planetary envelope ($\xi\,=\,0$), the MT would be dynamically stable as long as $\gamma\,\leq\,0.8$. 
We test different values of $\gamma$ with MESA and find that, depending on the planetary core mass, $M_{\rm c}$, the computation becomes numerically difficult for $\gamma$ values higher than about $0.6-0.8$. This suggests that the MT may indeed become dynamically unstable and that, according to the criterion in Eqn.~(\ref{eq:MTstability}), $\xi$ is inferred to be close to zero. To avoid the instability region, while still considering somewhat significant angular momentum loss from the system, we provide examples for non-conservative MT with $\gamma\,=\,0.5$ and $\gamma\,=\,0.6$ for the 1\,$M_{\oplus}$ and 5\,$M_{\oplus}$ core cases, respectively, and $\gamma\,=\,0.7$ for the higher core masses. 

 %{\bon Finally, for stellar wind mass loss we follow the test\_suite \emph{1M\_pre\_ms\_to\_wd} provided with MESA for the evolution of a 1$M_{\odot}$ star (\citealt{Bloecker95,Reimers75}). Similarly to Eqn.~(\ref{eq:JdotPE}), this channel of mass loss affects the orbital angular momentum $J_{\rm orb}$ according to: 
 %\begin{align}
%\left(\frac{\dot{J}_{\rm orb}}{J_{\rm orb}}\right)_{*wind}&=\frac{\dot{M}_{\rm *,wind}}{M_{\rm *}}\frac{q}{1+q}, 
%\label{eq:JdotWind}
%\end{align}
 %}.

%%%%%%%%%%%%%%%%%%%%%%%%%%
\newpage
\section{Results}\label{Example}
%%%%%%%%%%%%%%%%%%%%%%%%%%

We consider a typical hot-Jupiter system comprising a 1$M_{\odot}$ star and a 1$M_{\rm J}$ planet at solar metallicity. For the planet we took $M_{\rm c}\,=\,1\,M_{\oplus}, 5\,M_{\oplus}, 10\,M_{\oplus}, 15\,M_{\oplus}$ and, 30$\,M_{\oplus}$.  The initial period was set to $P_{\rm orb}=0.7$\,d in all cases. This corresponds to the orbital period at which a hot Jupiter with a $1.5R_{\rm J}$ radius would be at its Roche limit and it is chosen arbitrarily to have all systems starting with the same initial period ``close enough'' to the Roche limit. 

For the initial systems' age we chose 2\,Gyr ($t\,\simeq\,20\%\,t_{\rm MS}$), as it is at the low end of the ages of the currently known hot Jupiters closest to their Roche-limit separation (see Table~2 in \citealt{ValsecchiR2014b} and references therein; see also Section~\ref{discussion}).
Accordingly, we create 2\,Gyr old stellar and planetary models with MESA's ``single-star'' module. Each planet is irradiated according to the host star properties at 2\,Gyr and the 0.7\,d period ($F_{*}\,=\,5.4\,\times\,10^{9}$\,erg\,cm$^{-2}$\,s\,$^{-1}$).
%\footnote{\ron Isn't this gigantically larger than the threshold for radiation-recombination limited PE loss?}. 

The models are then used in MESA's ``binary'' module to compute the orbital evolution. During this step both stellar and planetary evolution, as well as irradiation and photo-evaporation are computed self-consistently (Section~\ref{Input Physics}). For the stellar spin, we choose an initial value of $\Omega_{*}\sim0.1\,\Omega_{\rm o}$, where $\Omega_{\rm o}$ is the orbital frequency. This is consistent with the observed slow stellar rotation rates for the tightest hot-Jupiter systems known ($\Omega_{*}\simeq0.1-0.2\,\Omega_{\rm o}$; see Table~1 in \citealt{ValsecchiR2014b}). 
Finally, as described in Section~\ref{RLO}, we consider both conservative and non-conservative MT. 

The results are presented as follows. In Section~\ref{Detailed Examples} we describe in detail the evolution of the 5\,$M_{\oplus}$ and 30\,$M_{\oplus}$ planetary core cases. We take these as extreme examples among those considered here. In fact, as described below (Section~\ref{The General Behavior for All Core Masses}), the behavior of the 1\,$M_{\oplus}$ core model needs a more in-depth investigation. For $M_{\rm c}=5\,M_{\oplus}$ and $M_{\rm c}=30\,M_{\oplus}$, we mainly focus on the non-conservative MT evolution and briefly describe the conservative case at the end of each section. 
For these same core masses, we also investigate the effect of varying the column density for irradiation in Section~\ref{varing sigma}, and the photo-evaporation recipe in Section~\ref{Varying Photo-Evaporation}. In Section~\ref{The General Behavior for All Core Masses} we present an overview of how the orbital period evolves as the planet loses mass, as well as how the planetary radius evolves with mass, for the full range of core masses. A summary of  the results for all core masses and physical assumptions is presented in Table~\ref{Tab:SummaryResults}. In all examples considered, the evolution is quite rapid after the planet is left with only a few percent of the envelope mass. Thus, we discuss the final stages of the planetary evolution qualitatively, guided by the relevant timescales entering the problem. 
%%%%%%%%%%%%%%%%%%%%%%%%%%%%%%%%%%%%%%%%%%%%%%%%
\begin{figure*} 
\epsscale{0.8}
\plotone{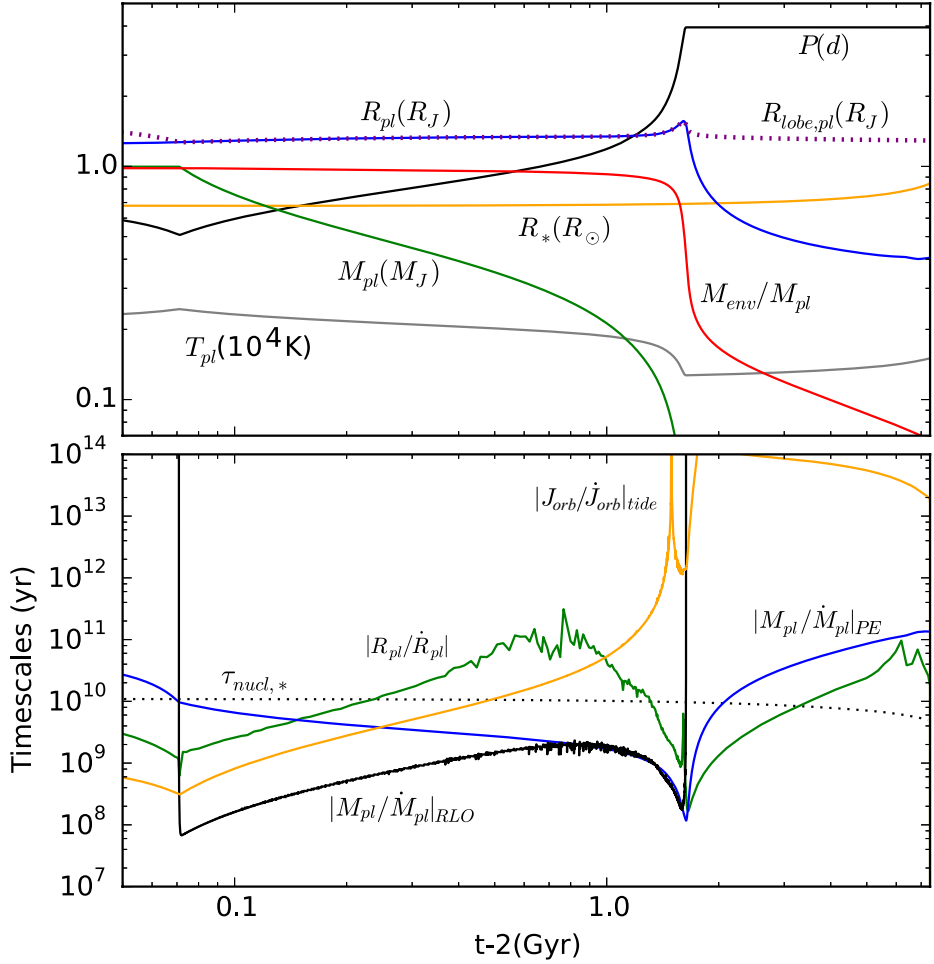}
\caption{Detailed evolution of some of the components and orbital parameters (top) and of the relevant timescales (bottom)  for a Jupiter with $M_{\rm c}\,=\,5\,M_{\oplus}$. For this evolution we adopt the parameters $\delta = 1$ and $\gamma = 0.6$. The subscripts ``RLO'', ``PE'', and ``Tides'' refer to the timescales associated with mass loss due to Roche-lobe overflow and photo-evaporation, and to tidal decay, respectively. For the latter we note that only stellar tides play a significant role, as the planetary tides timescale is always longer than a Hubble time.  The peaks in the tidal timescales occur when the contributions from stellar and planetary tides cancel out. The dotted blue curve in the upper panel indicates the Roche-lobe radius, while the dotted curve in the bottom panel is the stellar nuclear timescale. The RLO phase ends after about 1.6 \,Gyr, when the system is about 3.6\,Gyr old. For clarity, the timescale for the evolution of the planetary radius (bottom panel, green line) is computed taking the median of 20 consecutive values. The calculation ends when $M_{\rm env}/M_{\rm pl}\,\lesssim\,7$\% because of convergence problems.}
\label{MT_evolution_5Me_sigma1gcmM2_delta1_gamma0p6_2Gyr_Ftid50_epsilon0p1_Rxuv0p2percent_separateJdot_ls_MB} 
\end{figure*}

%%%%%%%%%%%%%%%%%%%%%%%%%%%%%%%%%%%%%%%%%%%%%%%%
\subsection{Detailed Examples: Different Core Masses}\label{Detailed Examples}
%%%%%%%%%%%%%%%%%%%%%%%%%%%%%%%%%%%%%%%%%%%%%%%%
\subsubsection{The $5\,M_{\oplus}$ Core Model}\label{Detailed Examples non-conservative 5}

\begin{figure} [!h]
\epsscale{1.15}
\plotone{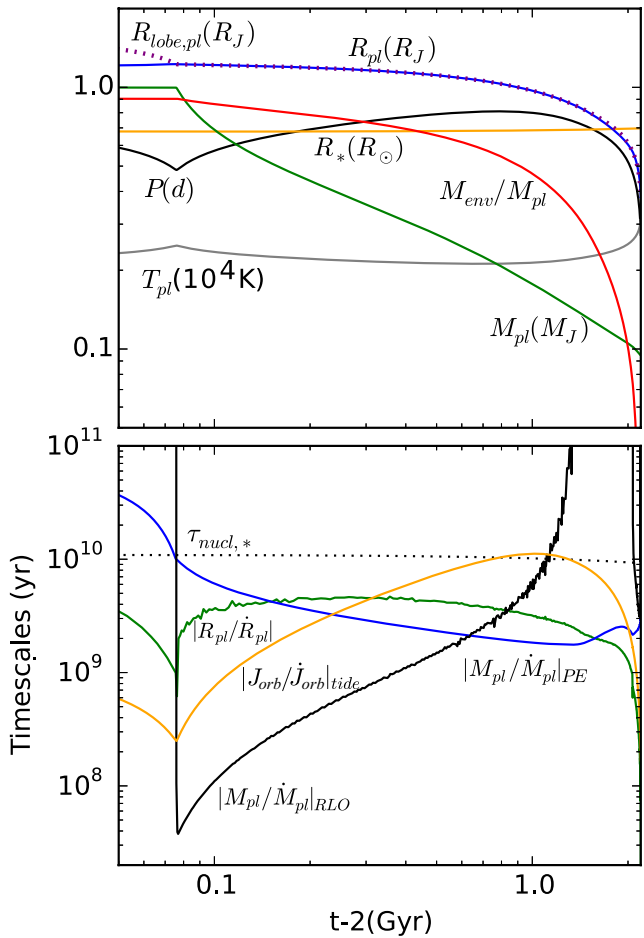}
\caption{Same as Figure~\ref{MT_evolution_5Me_sigma1gcmM2_delta1_gamma0p6_2Gyr_Ftid50_epsilon0p1_Rxuv0p2percent_separateJdot_ls_MB} but for a Jupiter with $M_{\rm c}\,=\,30\,M_{\oplus}$ and $\gamma\,=\,0.7$.  For clarity, $|M_{\rm pl}/\dot{M_{\rm pl}}|_{\rm RLO}$ is computed by taking the median of 10 consecutive values.}
\label{fig:MT_evolution_30Me_sigma1gcmM2_delta1_gamma0p7_2Gyr_Ftid50_epsilon0p1_Rxuv0p2percent_separateJdot_ls_MB} 
\end{figure}

%%%%%%%%%%%%%%%%%%%%%%%%%%%%%%%%%%%%%%%%%%%%%%%%
Figure~\ref{MT_evolution_5Me_sigma1gcmM2_delta1_gamma0p6_2Gyr_Ftid50_epsilon0p1_Rxuv0p2percent_separateJdot_ls_MB} shows the evolution of a Jupiter with a 5\,$M_{\oplus}$ core undergoing non-conservative MT. The top panel displays the evolution of various system and planetary properties, while the bottom panel investigates a number of different timescales of the system.  For convenience in studying the plot we arbitrarily subtract 2 Gyr from the time axis (this is just the time over which we evolved the star and the planet before inserting them into the binary evolution version of the MESA code; see discussion above).  Hereafter, when describing various features in Figure~\ref{MT_evolution_5Me_sigma1gcmM2_delta1_gamma0p6_2Gyr_Ftid50_epsilon0p1_Rxuv0p2percent_separateJdot_ls_MB} (as well as in Figure~\ref{fig:MT_evolution_30Me_sigma1gcmM2_delta1_gamma0p7_2Gyr_Ftid50_epsilon0p1_Rxuv0p2percent_separateJdot_ls_MB}) we refer to the time marked on the axis, i.e., after subtracting off a 2 Gyr reference time.
 
For the first $\sim\,$70\,Myr the planet underfills its Roche lobe while the orbit shrinks due to the tides transferring angular momentum from the orbit to the stellar spin, in a vain attempt to try to spin-up the star. Eventually, when $P_{\rm orb}$ shrinks to $\simeq\,0.5$\,d ($a\,\simeq\,0.01\,$AU) the planet fills its Roche-lobe.  This occurrence can be seen when the dotted curve in Fig.~\ref{MT_evolution_5Me_sigma1gcmM2_delta1_gamma0p6_2Gyr_Ftid50_epsilon0p1_Rxuv0p2percent_separateJdot_ls_MB} (indicating the Roche lobe radius) merges with the planetary radius curve (solid blue).  

The system remains in Roche lobe contact for the ensuing 1.6 Gyr, while the planet loses more than 95\% of its mass.  Concomitantly, the orbital period grows from about 0.5 to 3.9 days.  During this same interval, the effective temperature of the planet decreases from about 2500 K to about 1300 K, due largely to the increasing distance between the host star and the planet.

During the RLO phase there is a competition between the tidal effects which tend to drive orbital decay and mass transfer which generally drives orbital expansion [see Eqn.~(\ref{eq:aRLO})].  Both effects operate simultaneously, and the one that dominates depends on the response of the planet's radius to mass loss.  For Roche-lobe overflowing objects with mass lower than the accreting star, there is a universal relation between the orbital period, and mass and radius of the donor [see, e.g., Eqn.~(2) of \citealt{HNR01}]:
\begin{equation}
P_{\rm orb} \simeq 0.4 \,(M_{\rm pl}/M_J)^{-1/2} (R_{\rm pl}/R_J)^{3/2} ~{\rm days}
\label{Eqn:PRM}
\end{equation}
Thus, if the mass transfer remains stable, then the orbital period will grow or shrink in accordance with how the planetary radius changes with mass loss.  In this example with a 5 $M_\oplus$ core, we can see from Fig.~\ref{MT_evolution_5Me_sigma1gcmM2_delta1_gamma0p6_2Gyr_Ftid50_epsilon0p1_Rxuv0p2percent_separateJdot_ls_MB} that the planet's radius does not vary very much during most of the RLO phase, and in fact, slightly increases to 1.6 $R_J$.  Therefore, $P_{\rm orb}$ will simply grow mostly as $\propto M_{\rm pl}^{-1/2}$.  By the end of the RLO phase the planet's mass has shrunk to $8.5 \, M_\oplus$, and, according to Eqn.~(\ref{Eqn:PRM}) $P_{\rm orb}$ should be about 4 days, which it is.  The Roche lobe overflow phase ends when the combination of tidal decay of the orbit and planetary expansion is no longer sufficient to keep the planet filling its Roche lobe, and the planet, which is now quite low in mass, starts to shrink well inside its Roche lobe.

The post-RLO phase is driven mainly by photo-evaporation. In fact, the tidal decay timescale becomes longer than a Hubble time for the majority of this phase due to the longer orbital period and the greatly reduced planetary mass. Photo-evaporation removes mass from the planet at a rate of about $10^{-16}-\,10^{-15}M_{\odot}\,$yr$^{-1}$ ($10^{10}-10^{11}\,$g~s$^{-1}$) and it does not affect the orbital separation significantly [see Eqn.~(\ref{eq:a_dotPE})]. For the 5.4\,Gyr duration of the post-RLO phase the planet remains in a 3.9\,d orbit and its mass decreases from about 8.5\,$M_{\oplus}$ to about 5.4\,$M_{\oplus}$ (when $M_{\rm env}/M_{\rm pl}\,\simeq\,$7\%), when the calculation ends because of convergence problems.  

The bottom panel of Fig.~\ref{MT_evolution_5Me_sigma1gcmM2_delta1_gamma0p6_2Gyr_Ftid50_epsilon0p1_Rxuv0p2percent_separateJdot_ls_MB} displays various timescales for this evolution, including the mass loss timescales via the photo-evaporative wind, $|M_{\rm pl}/\dot{M}_{\rm pl}|_{\rm PE}$, and due to RLO, $|M_{\rm pl}/\dot{M}_{\rm pl}|_{\rm RLO}$, as well as the timescale for tidal decay of the orbit, $|J_{\rm orb}/\dot{J}_{\rm orb}|_{\rm tides}$, and for the thermal expansion/contraction of the planetary radius, $|R_{\rm pl}/\dot{R}_{\rm pl}|$.  What we see is that, after RLO commences, the black curve, showing the mass transfer timescale (due to RLO), lies a factor of about 8 lower than the tidal decay timescale, and much lower than the mass loss timescale associated with PE winds. We learn from Eqn.~(\ref{eq:MdotPlanet}) that the mass loss rate due to RLO is therefore essentially proportional to the rate of decay of the orbit due to tides, and inversely proportional to the denominator which is $\xi/2 - \gamma \delta +5/6 = \xi/2 +0.23$.  Putting these together implies that the denominator must equal approximately ~1/8.  From this we can infer that the effective adiabatic index of the planet during most of the RLO phase is $\xi \simeq -0.2$, in other words the planet reacts to adiabatic mass loss by slightly expanding.  Later in the RLO phase, the tidal decay timescale greatly increases due to the increasing orbital separation, but the thermal expansion timescale of the planet decreases dramatically to pick up the slack of the declining tidal effects.  We also see that the mass loss rate in a PE driven wind dramatically increases (i.e., the timescale decreases) due to the increasing radius and the decreasing mass of the planet [see Eqn.~(\ref{eq:MdotPE_eLim})]. 

Even though our evolution calculation ends when the envelope mass fraction drops to about 7\%, we argue that eventually, the host star will approach the end of its main sequence lifetime and it will begin expanding. The increase in $R_{*}$ will cause a concomitant increase in both the tidal decay rate and in the amount of irradiation received by the planet. The latter occurs because $R_{*}$ evolves faster than $T_{*}$ in Eqn.~(\ref{eq:Fstar}). Such an increase in irradiation will yield an increase in $R_{\rm pl}$ and, as a consequence, even stronger photo-evaporation. 
%The rate of mass loss from the planet increases by about an order of magnitude ($\sim\,10^{-14}\,M_{\odot}\,$yr$^{-1}$ = $10^{12}\,$g~s$^{-1}$).
What happens to this system after the envelope of the planet has been removed, leaving only its rocky core, depends mainly on the tidal timescale compared with the stellar nuclear evolution timescale. The former determines the rate of tidal decay (leading to a second planetary RLO), while the latter determines the rate of stellar expansion (leading to stellar RLO). Given the (very steep) dependence of the tidal decay timescale as $(a/R_{*})^{8}$ (e.g., \citealt{ValsecchiR2014}), we argue that tides will likely drive the planet down to its Roche limit. 
Here, it will be stripped of the remaining envelope and, assuming a constant density for the core, it will be consumed at approximately constant $P_{\rm orb}$ on the rapid tidal decay timescale. This follows from \citeauthor{Paczynski71}'s (\citeyear{Paczynski71}) approximation for the Roche limit separation $a_{\rm R}\,=\,(R_{\rm pl}/0.462) (M_{*}/M_{\rm pl})^{1/3}\propto (M_{*}/\rho_{\rm pl})^{1/3}$, where $\rho_{\rm pl}$ denotes the mean density of the planet.

%%%%%%%%%%%%%%%%%%%%%%%%%%%%%%%%%%%%%%%%%%%%%%%%
%%%%%%%%%%%%%%%%%%%%%%%%%%%%%%%%%%%%%%%%%%%%%%%%

The above discussion was for the case of non-conservative MT (with $\gamma =0.6$).  In the case of conservative mass transfer, we find that the orbital evolution proceeds similarly to the examples presented above, but the duration of the various phases is different. 
In fact, when MT is conservative, the tidally-driven orbital decay is counteracted by a more substantial RLO-driven orbital expansion [$\delta\,=\,0$ in Eqn.~(\ref{eq:aRLO})], which results in a {\em slower} overall evolution.
This is summarized in Table~\ref{Tab:SummaryResults}. In particular, the RLO phase lasts for about 4.5\,Gyr, leaving an 8.3\,$M_{\oplus}$ planet in a 3.4\,d orbit. The remainder of the evolution proceeds similarly to the non-conservative case. However, the evolution when little envelope mass is left is shorter. This is due to the irradiation-driven decrease in planetary mass loss timescale ($|M_{\rm pl}/\dot{M}_{\rm pl, PE}|$) due to the star approaching the end of its main sequence. 

%%%%%%%%%%%%%%%%%%%%%%%%%%%%%%%%%%%%%%%%%%%%%%%%
\subsubsection{The $30\,M_{\oplus}$ Core Model}\label{Detailed Examples non-conservative 30}
%%%%%%%%%%%%%%%%%%%%%%%%%%%%%%%%%%%%%%%%%%%%%%%%

Figure~\ref{fig:MT_evolution_30Me_sigma1gcmM2_delta1_gamma0p7_2Gyr_Ftid50_epsilon0p1_Rxuv0p2percent_separateJdot_ls_MB} shows the evolution of a Jupiter with a 30\,$M_{\oplus}$ core undergoing non-conservative MT (with $\gamma = 0.7$).  As in Fig.~\ref{MT_evolution_5Me_sigma1gcmM2_delta1_gamma0p6_2Gyr_Ftid50_epsilon0p1_Rxuv0p2percent_separateJdot_ls_MB} (for the case of a 5 $M_\oplus$ core), the top panel presents the evolution of various system and planetary properties, while the bottom panel displays a number of different timescales of the system.
For the first $\sim\,$70\,Myr tides cause the orbit to shrink and the planet fills its Roche lobe when $P_{\rm orb}\,\simeq\,0.5$\,d (the same as for the 5 $M_\oplus$ core case).  
 
At the onset of RLO, the orbit expands as mass is removed from the planet at a rate of $\sim\,10^{-13}-10^{-12}\,M_{\odot}\,$yr$^{-1}$ ($10^{13}-10^{14}\,$g~s$^{-1}$).  However, after about a Gyr, the orbit begins shrinking once the period has grown to only $\sim$0.8 days. 
Overall, we follow the RLO phase for about 2.1\,Gyr, to the point where the planetary envelope has been completely removed. At this time, the orbital period has shrunk to 0.3\,d, and the effective temperature of the planet has increased to nearly 3000\,K. 

The difference in evolutionary history between this planet with a 30 $M_\oplus$ core and the one with a 5\,$M_{\oplus}$ core (described in Section~\ref{Detailed Examples non-conservative 5}) results largely from the fact that planets with more massive cores exhibit an earlier (i.e., at higher $M_{\rm pl}$; Figure ~\ref{fig:Mpl_vs_Porb_sigma1gcmM2_multi_2Gyr_Ftid50epsilon0p1Rxuv0p2percent_separateJdot_ls_MB}) decrease in planetary radius with continuing mass loss.  During the RLO phase with a 30 $M_\oplus$ core, the mass of the planet decays essentially as a power law in time, while the radius does not decrease significantly until $M_{\rm pl}$ drops below $\sim$60 $M_\oplus$.  From Eqn.~(\ref{Eqn:PRM}) we can deduce that this combination of mass and radius changes will lead to a steady increase in the orbital period. However, once the radius starts to decline substantially, the $R_{\rm pl}^{3/2}$ dependence in Eqn.~(\ref{Eqn:PRM}) dominates the orbital period evolution, and $P_{\rm orb}$ starts to decay. %This behavior of the radius with decreasing mass yields a less severe mass loss during RLO and a consequent less dramatic orbital expansion. 
%As a result, the orbital evolution proceeds at shorter orbital periods and the planet never shrinks back inside its Roche-lobe. 

We can gain some further insight into the evolution of this system by considering the timescales displayed in the bottom panel of Figure~\ref{fig:MT_evolution_30Me_sigma1gcmM2_delta1_gamma0p7_2Gyr_Ftid50_epsilon0p1_Rxuv0p2percent_separateJdot_ls_MB}.  For most of the evolution, the tidal decay timescale ($|J_{\rm orb}/\dot{J}_{\rm orb}|_{\rm tides}$) is about 6 times longer than the mass loss timescale ($|M_{\rm pl}/\dot{M}_{\rm pl}|_{\rm RLO}$). Furthermore, at least for the earlier portion of the RLO phase, both the contraction timescale of the planet ($|R_{\rm pl}/\dot{R}_{\rm pl}|$) and the timescale of photo-evaporation ($|M_{\rm pl}/\dot{M}_{\rm pl}|_{\rm PE}$) are both longer yet than the tidal decay timescale.  From this we can infer that the denominator in Eqn.~(\ref{eq:MdotPlanet}) must be $\simeq 1/6$.  In turn, we can conclude that $\xi_{\rm ad} \simeq 0.07$ (i.e., very close to zero).  During this earlier portion of the RLO evolution, the orbital period grows and the mass of the planet declines, a combination that leads to an ever increasing tidal decay timescale (i.e., weakening of the tidal evolution of the orbit).  

During the later portion of the RLO phase, both $|R_{\rm pl}/\dot{R}_{\rm pl}|$ and $|M_{\rm pl}/\dot{M}_{\rm pl}|_{\rm PE}$ become {\em shorter} than the tidal driving timescale.  This means that there is close competition in the numerator of Eqn.~({\ref{eq:MdotPlanet}) for maintaining Roche-lobe contact among all three terms: (i) tidal decay; (ii) the photo-evaporative mass loss term (tending to shrink the Roche-lobe radius of the planet and maintain RLO), and (iii) the shrinkage of the planet as it loses mass, which tends to push the planet back within its Roche-lobe.  Apparently, the combination is sufficient to maintain RLO until we terminate the evolution.  

Our calculation ends when the planetary envelope has been completely removed. However, the core itself may well eventually undergo RLO and, if the core has a constant density, it will be consumed at constant orbital period on the rapid tidal decay timescale ($\lesssim\,100\,$Myr, at the end of the evolution in Figure~\ref{fig:MT_evolution_30Me_sigma1gcmM2_delta1_gamma0p7_2Gyr_Ftid50_epsilon0p1_Rxuv0p2percent_separateJdot_ls_MB}).

The conservative mass transfer case for the same model with a 30 $M_\oplus$ core differs only insofar as the duration of the RLO phase is concerned (see Table~\ref{Tab:SummaryResults}). This is the same as what we found for the conservative vs.~non-conservative cases with a 5$\,M_{\oplus}$ core.
In fact, during conservative MT, the RLO phase until the envelope is removed lasts about 3.8\,Gyr. At the end of the calculation $P_{\rm orb}\,=\,0.3$\,d.

%%%%%%%%%%%%%%%%%%%%%%%%%%%%%%%%%%%%%%%%%%%%%%%%
\subsubsection{Varying the Column Density for Irradiation}\label{varing sigma}
%%%%%%%%%%%%%%%%%%%%%%%%%%%%%%%%%%%%%%%%%%%%%%%%

For highly irradiated planets with $M_{\rm pl} < M_{\rm J}$, the quasi-equilibrium mass-radius relations of \cite{FortneyMB07} in Figure~\ref{fig:mass_radius_compareFortney_irradiation_radiusAt1bar} are bracketed by those computed with MESA for irradiation absorption column densities of $\Sigma_{\rm pl}$\,=\,(1$-$100)\,g/cm$^{2}$. We tested the effect on planetary RLO evolution by increasing $\Sigma_{\rm pl}$ by two orders of magnitude in the 5$\,M_{\oplus}$ and 30$\,M_{\oplus}$ core mass models. We find that the overall evolution does not change significantly (see Table~\ref{Tab:SummaryResults}). 
Qualitatively, for the 5\,$M_{\oplus}$ core case, at the beginning of the calculation the radius at higher $\Sigma_{\rm pl}$ is a few percent larger. As a result, RLO starts at a longer orbital period and, for the same evolutionary time, it continues at longer $P_{\rm orb}$. When the planet retreats back within its Roche-lobe, the longer orbital period yields a less severe mass loss via photo-evaporation. Consequently, the planet retains a higher envelope mass fraction ($f$ in Table~\ref{Tab:SummaryResults}) for a longer time.
%Qualitatively, $R_{\rm pl}$ is a few percent larger {\bf for most of the evolution}. As a result, for the same {\bf envelope mass fraction}, RLO starts and proceeds at a longer orbital period. For the 5\,$M_{\oplus}$ core model, the orbital period is a few percent longer than in the smaller $\Sigma_{\rm pl}$ case. 
A similar behavior occurs for the 30$\,M_{\oplus}$ core model, but the percent difference in radius (orbital period) increases from $\sim\,5\%$ to $\sim\,15\%$  (from $\sim\,5\%$ to $\sim\,20\%$) between the beginning and the end of the calculation. Again, mass loss proceeds on a longer timescale and, for the same evolutionary time, the models with higher $\Sigma_{\rm pl}$ retain a larger fraction of the envelope for longer. 

\subsubsection{Varying Photo-Evaporation}\label{Varying Photo-Evaporation}
%%%%%%%%%%%%%%%%%%%%%%%%%%%%%%%%%%%%%%%%%%%%%%%%

Formally, the prescription of \cite{Murray-Clay+09} uses $F_{\rm XUV}\sim10^{4}$\,erg\,cm$^{-2}$\,s$^{-1}$ as a threshold between the energy limited and radiation/recombination limited regimes. Below we investigate whether our results for the 5\,$M_{\oplus}$ and 30\,$M_{\oplus}$ core cases change significantly by using  Eqn.~(\ref{eq:MdotPE_rrLim}) when $F_{\rm XUV}>10^{4}$\,erg\,cm$^{-2}$\,s$^{-1}$ and Eqn.~(\ref{eq:MdotPE_eLim}) otherwise. The results are summarized in Table~\ref{Tab:SummaryResults} (denoted with ``e-Lim + rr-Lim'' in the column named ``PE''). 

The planetary mass loss rate in the radiation/recombination limited regime is slower. This naturally leads to a longer duration of the RLO phase for all examples considered. For the 5\,$M_{\oplus}$ core undergoing conservative MT, $R_{\rm pl}$ is a few percent larger. As a result, the orbit evolves at longer period. However, because of the longer RLO phase, by the time the planet is left with about 10\% of its envelope mass the star is approaching the end of its main sequence. Consequently, the increase in $R_{\rm pl}$ driven by the increase in stellar irradiation leads to faster mass loss via photo-evaporation towards the end of the calculation. 
For the 5\,$M_{\oplus}$ core undergoing non-conservative MT, the evolution follows this same line of logic, but the planetary radius is a few percent smaller for most of the evolution. As a result, the orbit evolves at shorter period. For the 30$M_{\oplus}$ core case, the response of $R_{\rm pl}$ to mass loss for the different photo-evaporation prescriptions differs by only $1-2$\%.
% {\ron (differs from what?)}. 
Therefore, only the duration of the RLO phase changes significantly, while the orbital period at the various stages of Table~\ref{Tab:SummaryResults} is not significantly affected.

In summary, none of our basic results would change significantly if we were to utilize both the energy-limited and radiation/recombination-limited prescriptions for photo-evaporation over the entire evolution.  However, certainly a number of the details of the evolutionary models would look somewhat different.

\begin{figure} [!h]
\epsscale{1.2}
\plotone{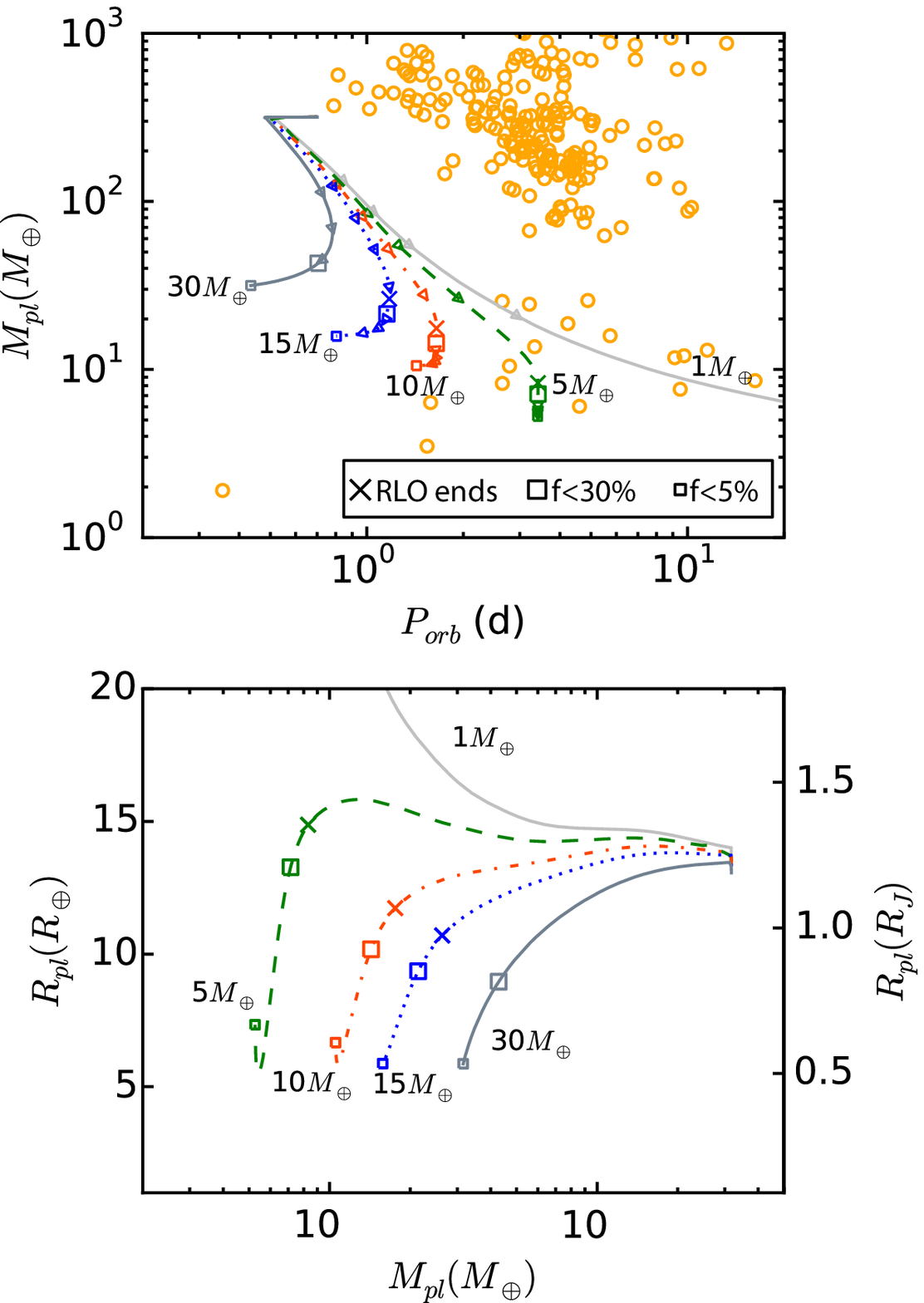}
\caption{Planetary mass as a function of the orbital period (top) and mass-radius relation (bottom) for conservative MT (with $\delta = 0$). Because some of the evolutionary calculations become numerically challenging when the mass fraction in the envelope drops below a few percent, we show the evolution up to when $M_{\rm env}/M_{\rm pl}$ drops just below 5\%. The open orange circles are confirmed exoplanetary systems (NASA Exoplanet Archive, 13 January 2015) with observationally inferred $M_{\rm pl}$ and $P_{\rm orb}$, and hosting one planet only, for simplicity. 
The system GJ 436b~\citep{ButlerGJ436b2004} would be located at $P_{\rm orb} \simeq 2.6$\,d and $M_{\rm pl}\simeq$22$M_{\oplus}$.
The colored lines are our evolutionary models for different core masses.
%$M_{\rm c}$\,=\,30$\,M_{\oplus}$ (grey), 15$\,M_{\oplus}$ (blue), 10$\,M_{\oplus}$ (red), 5$\,M_{\oplus}$ (green), and 1$\,M_{\oplus}$ (light grey). 
The colored arrows along each evolutionary track in the top panel mark 1\,Gyr intervals and denote time evolution. For the remaining symbols:``$\times$s'' mark the end of RLO, ``$\square$s'' mark times when the mass in the envelope drops below $f = $ 30\%, and 5\% of the total mass. For the 5\,$M_{\oplus}$ core model, the various symbols in the top panel are all superimposed at the end of the evolution, where the planet spends a few Gyrs.}
\label{fig:Mpl_vs_Porb_sigma1gcmM2_delta0_2Gyr_Ftid50epsilon0p1Rxuv0p2percent_separateJdot_ls_MB} 
\end{figure}
%%%%%%%%%%%%%%%%%%%%%%%%%%%%%%%%%%%%%%%%%%%%%%%%
\begin{figure}
\epsscale{1.2}
\plotone{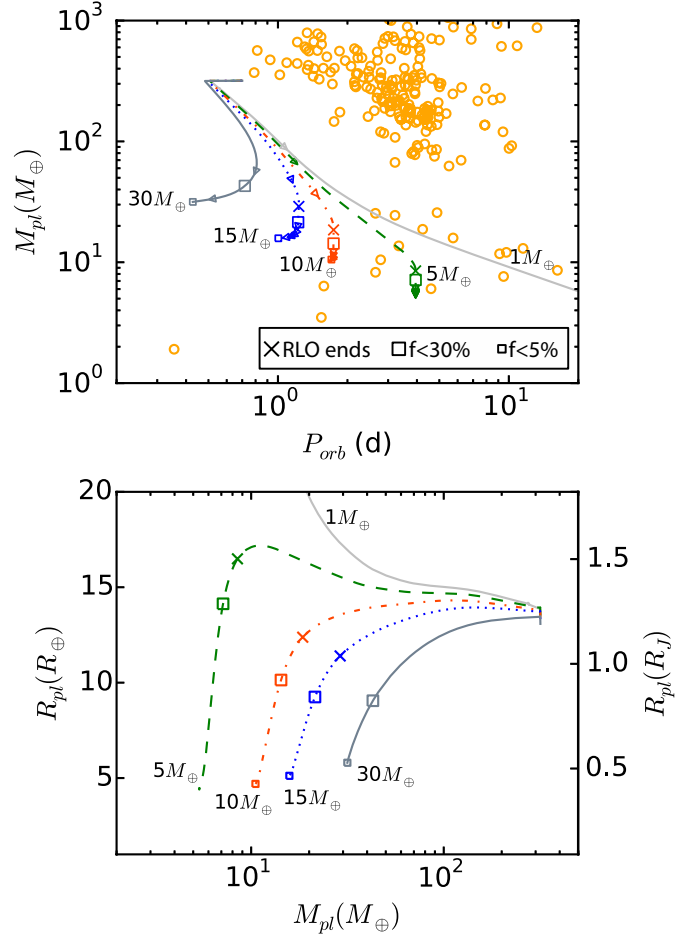}
\caption{Same as Figure~\ref{fig:Mpl_vs_Porb_sigma1gcmM2_delta0_2Gyr_Ftid50epsilon0p1Rxuv0p2percent_separateJdot_ls_MB} but for non-conservative MT (with $\delta = 1$). We use $\gamma\,=\,$0.5, 0.6, and 0.7 for $M_{\rm c}\,=\,1\,M_{\oplus}$, 5\,$M_{\oplus}$, and $\geq\,10\,M_{\oplus}$, respectively. Note, for the 5\,$M_{\oplus}$ core model the evolution terminates because of convergence problems when the envelope contains $\simeq\,6$\,\%  of the total mass. For the 5\,$M_{\oplus}$ core model described in detail in the text, the various symbols in the top panel are all superimposed at the end of the evolution, where the planet spends few Gyrs.}
\label{fig:Mpl_vs_Porb_sigma1gcmM2_multi_2Gyr_Ftid50epsilon0p1Rxuv0p2percent_separateJdot_ls_MB} 
\end{figure}
%%%%%%%%%%%%%%%%%%%%%%%%%%%%%%%%%%%%%%%%%%%%%%%%

\subsection{A Range of Evolutionary Models}\label{The General Behavior for All Core Masses}
A range of evolutionary models covering different core masses for both conservative and non-conservative MT are presented here. The results are summarized in Figures~\ref{fig:Mpl_vs_Porb_sigma1gcmM2_delta0_2Gyr_Ftid50epsilon0p1Rxuv0p2percent_separateJdot_ls_MB} and \ref{fig:Mpl_vs_Porb_sigma1gcmM2_multi_2Gyr_Ftid50epsilon0p1Rxuv0p2percent_separateJdot_ls_MB}, respectively. The top panels show the evolution of the orbital period with the planet's mass.  For comparison, the positions of the observed planets are the orange open circles. The bottom panels show the corresponding evolutionary tracks for planetary mass and radius. The evolution tracks for different core masses are denoted with different colors and line-styles. For all cases considered, the MT proceeds on a timescale longer than the thermal timescale of the planet. Thus, the planet remains in near thermal equilibrium throughout the MT.  Here we note that our 1\,$M_{\oplus}$ core model shows a severe increase in $R_{\rm pl}$ that needs further investigation. In fact, $R_{\rm pl}$ increases up to about $10\,R_{\rm J}$ when $M_{\rm pl}\,\simeq\,1\,M_{\oplus}$. However, we note also that the core-less models in Figure~8 of \cite{FortneyMB07} have radii ranging up to $R_{\rm pl}\simeq 2.3\,R_{\rm J}$ when $M_{\rm pl}\,\sim\,30\,M_{\oplus}$ (the lowest mass considered by \citealt{FortneyMB07} for a core-less Jupiter), depending on the age of the planet and the level of irradiation. This is consistent with our radii for masses down to $\sim$$10\,M_{\oplus}$. We omit the 1$\,M_{\oplus}$ core model from the subsequent discussion.

The evolution in time in the $M_{\rm pl}-P_{\rm orb}$  plane is downward (i.e., shrinking mass), initially toward longer periods, but then decaying toward shorter $P_{\rm orb}$. The tracks in the ($R_{\rm pl}-M_{\rm pl}$) diagram proceed from the right to the left. The shape of the evolutionary tracks in such diagrams depends on the competing effects of stellar tides (tending to shrink the orbit) and RLO (tending to expand the orbit). As described in Section~\ref{Detailed Examples non-conservative 5}, the $R_{\rm pl}-M_{\rm pl}$ tracks in the bottom panels of Figures~\ref{fig:Mpl_vs_Porb_sigma1gcmM2_delta0_2Gyr_Ftid50epsilon0p1Rxuv0p2percent_separateJdot_ls_MB} and \ref{fig:Mpl_vs_Porb_sigma1gcmM2_multi_2Gyr_Ftid50epsilon0p1Rxuv0p2percent_separateJdot_ls_MB} partly determine which mechanism ends up dominating in terms of net orbital contraction or expansion [see Eqn.~(\ref{Eqn:PRM})]}. 
%In fact, in addition to tides, RLO affects the orbital separation according to Equation(\ref{eq:aRLO}), and $|\dot{M}_{\rm pl,RLO}/M_{\rm pl}|$ is partly determined by the response of the planetary radius to mass loss.

At the onset of RLO, the nearly constant or increasing $R_{\rm pl}$ with decreasing $M_{\rm pl}$ 
% (in particular, if $\xi < 1/3$) this should be deleted
causes the RLO term to dominate over the tidal contribution.  As a result the orbit expands. This behavior persists until the mass-radius relation begins steepening, becoming more positive. At this point, the tidal term becomes more significant than the RLO term and the orbit begins to shrink. 
In some cases ($M_{\rm c}\,=\,5\,M_{\oplus}, 10\,M_{\oplus}, 15\,M_{\oplus}$), the combination of the planet shrinking in response to mass loss and the weakening tidal forces with decreasing $M_{\rm pl}$ causes the RLO phase to terminate and the planet to detach (``$\times$'' symbols). In a few cases ($M_{\rm c}\,=\,5\,M_{\oplus}, 10\,M_{\oplus}$ during conservative MT), as the star approaches the end of its main sequence, the increasing amount of irradiation received by the planet causes $R_{\rm pl}$ to increase again (e.g., \citealt{FortneyNettelmann2010}). Note that none of the observed exoplanets in Figures~\ref{fig:Mpl_vs_Porb_sigma1gcmM2_delta0_2Gyr_Ftid50epsilon0p1Rxuv0p2percent_separateJdot_ls_MB} and ~\ref{fig:Mpl_vs_Porb_sigma1gcmM2_multi_2Gyr_Ftid50epsilon0p1Rxuv0p2percent_separateJdot_ls_MB} is currently in RLO. Therefore, observations should be compared with the portion of the evolutionary tracks where the planet is detached. There are some 6-7 observed systems shown in these figures which are in the vicinity of our evolution tracks, after Roche-lobe overflow has stopped, and our models may be directly applicable to them.  Note also that \cite{EhrenreichNature2015} recently reported the discovery of a large exospheric cloud surrounding the Neptune-mass exoplanet GJ 436~\citep{ButlerGJ436b2004}, composed mainly of hydrogen atoms. The average observed mass loss rate implies an efficiency for converting X-ray and extreme UV energy into mass loss of about 1\%. In Figures~\ref{fig:Mpl_vs_Porb_sigma1gcmM2_delta0_2Gyr_Ftid50epsilon0p1Rxuv0p2percent_separateJdot_ls_MB} and ~\ref{fig:Mpl_vs_Porb_sigma1gcmM2_multi_2Gyr_Ftid50epsilon0p1Rxuv0p2percent_separateJdot_ls_MB} this planet would be located at $P_{\rm orb} \simeq 2.6$\,d and $M_{\rm pl}\simeq$22$M_{\oplus}$.

Finally, we note that the shape of the evolutionary tracks does not change significantly in the $M_{\rm pl}-P_{\rm orb}$ diagram between conservative and non-conservative MT. However, as summarized in Table~\ref{Tab:SummaryResults} and explained in Section~\ref{Detailed Examples}, whether mass is lost from the system or not does affect the duration of the various phases mentioned above.  
%%%%%%%%%%%%%%%%%%%%%%%%%%%%%%%%%%%%%%%%%%%%%%%%
%%%%%%%%%%%%%%%%%%%%%%%%%%%%%%%%%
\section{Discussion} \label{discussion}
%NONE OF THIS BELONGS IN THE DISCUSSION - if anywhere, it belong in the conclusions.
%The stable MT calculations presented here show that the more non-conservative the MT, the shorter the duration of the RLO phase. In turn, the duration of the detached phase is correspondingly longer.  In particular we find that a hot Jupiter going through a phase of non-conservative MT detaches when its envelope contains tens of percent of the total mass (i.e., Figures~\ref{MT_evolution_5Me_sigma1gcmM2_delta1_gamma0p6_2Gyr_Ftid50_epsilon0p1_Rxuv0p2percent_separateJdot_ls_MB}, \ref{fig:Mpl_vs_Porb_sigma1gcmM2_delta0_2Gyr_Ftid50epsilon0p1Rxuv0p2percent_separateJdot_ls_MB}, and  \ref{fig:Mpl_vs_Porb_sigma1gcmM2_multi_2Gyr_Ftid50epsilon0p1Rxuv0p2percent_separateJdot_ls_MB}). After RLO has ended, the planet spends a few Gyr losing the remainder of its envelope via photo-evaporation at nearly constant orbital period. For example, a Jupiter with a 5$\,M_{\oplus}$ core (Section~\ref{Example}), remains detached for a few Gyrs at an orbital period of 3.9\,d. During this time, its envelope mass fraction drops from $\sim\,20$\% when the system is about 3.8\,Gyr to about 7\% when the system is about 8.8\,Gyr old, and the star approaches the end of its main sequence lifetime. A similar evolution occurs for $M_{\rm c}\,=\,$10$\,M_{\oplus}$ and 15$\,M_{\oplus}$, with an increasingly shorter duration of the detached phase for increasingly massive cores. Interestingly, the bulk density of \emph{known} 
Our calculations seem very promising for explaining some of the super-Earth and sub-Neptunes-type planets whose bulk density suggests that they consist of a core (rocky or icy) surrounded by a H/He envelope comprising up to tens of a percent of the total mass \citep{LopezFortney2014}. While this agreement with our MT models is very encouraging, our calculation neglects some important effects which we discuss below. Specifically, in Section~\ref{Mass Transfer Scenarios and Stability} we discuss possible MT scenarios, while in Section~\ref{Planetary Tides} we discuss our assumptions on planetary tides. A discussion of observational signatures is in Section~\ref{Obser_Signat}.
%and as we discuss below, observations suggest that super-Earth and sub-Neptunes-type planets may have tens of a percent of the total mass in their H/He envelopes.

\subsection{Mass Transfer Scenarios and Stability}\label{Mass Transfer Scenarios and Stability}  
In this work we neglected the effects of magnetic fields and stellar winds on the RLO material. These mechanisms can affect the flow of MT (e.g., \citealt{CohenGlocer12,OwenAdams2014}), potentially playing a crucial role in determining whether any mass is transferred or an accretion disk ever forms. 
On the opposite side of conservative MT, is the case where mass is blown away directly from the planet or the inner Lagrangian point ($L_{\rm 1}$). 
For a 1$M_{\rm J}$ planet and a 1\,$M_{\odot}$ star, $L_{\rm 1}$ is located at $\simeq\,0.93\,a$ [Eqn.~(10) of \citealt{LaiHvdH2010}]. In the formalism of Section~\ref{Orbital Evolution Model} these scenarios can be reproduced by setting $\delta\,=\,1$ and $\gamma\,=\,1$ (mass lost from the planet) or $\gamma\,=\,0.97$ (mass loss from $L_{\rm 1}$). In this configuration, Eqn.~(\ref{eq:MTstability}) requires $\xi$ to be larger than (0.27-0.33) in order for the MT to be dynamically stable. The values of $\xi$ inferred in Sections~\ref{Detailed Examples non-conservative 5} and \ref{Detailed Examples non-conservative 30} for our extreme core masses ($\xi = -0.2$ and 0.07 for the 5$M_{\oplus}$ and 30$M_{\oplus}$, respectively) suggest that the MT will likely be dynamically unstable if $\gamma \gtrsim 0.97$. Furthermore, we have performed test runs with MESA fixing $\delta=1$ while varying $\gamma$. We find that the MT may become dynamically unstable for $\gamma\,\geq\,0.8$. In fact, the computation becomes numerically difficult, with the integrator time-step dropping to less than one month. This limit on $\gamma$ for stability suggests that the effective value of $\xi$ is close to zero, or negative (as directly computed within MESA). %$\xi\simeq\,0$.

%In fact, at the onset of MT $R_{\rm pl}$ increases as $M_{\rm pl}$ decreases for  $M_{\rm c}\,<30\,M_{\oplus}$, and the decrease of  $R_{\rm pl}$ for the 30\,$M_{\oplus}$ core model is too shallow.
To gain better insight into more physically meaningful values of the parameter $\gamma$, consider the case 
%\footnote{\ron What is ``alternative'' about it?  That is the scenario we have been studying.}, 
where matter flows in a narrow stream from the $L_{\rm 1}$ point, and forms a ring around the star which is then blown away. 
This case has been discussed in the context of MT stability in stellar binary systems (e.g., \citealt{HutPaczynski1984,VerbuntRappaport1988}). If, in fact, 
%the ring size is too narrow to tidally interact with the donor, 
 the lifetime of the gas in the ring, before being blown away, is shorter than the viscous timescale for the ring, then the angular momentum may not be returned to the orbit. In this case, the angular momentum leaving the system would be determined by how much angular momentum a particle has before it is blown away (i.e., by the mean ring radius $r_{\rm d}$). 
We combine calculations of $r_{\rm d}$ by \citet{LubowShu1975} and \citet{HutPaczynski1984}, covering $M_{\rm pl}/M_{*}$ down to $10^{-3}$, with our own calculations, extending these down to $M_{\rm pl}/M_{*}\,=\,10^{-6}$. We find $r_{\rm d}$ to range between $\simeq~0.7-0.9$ for a 1\,$M_{\rm J}-10\,M_{\oplus}$ planet. 
%These imply values of $\gamma$ that range from 0.84 to 0.95.  
This range implies $\gamma$ values between $\simeq~0.84-0.95$ and a value of $\xi$ for stability between $\simeq$\,0$-$0.23 [Eqn.~(\ref{eq:MTstability})]. This scenario appears borderline between stable and unstable MT. Clearly any scenario where $\gamma\,>\,1$ would be dynamically unstable.

If the MT were indeed dynamically unstable, one could envision the system undergoing a common-envelope-like evolution (`CE'), that is somewhat distinct from the standard binary stellar evolution picture \citep{Webbink1984}.  We envision that the envelope of the planet would quickly flow though the inner Lagrange point and form a disk-like structure around the host star.  The core of the planet would then find itself orbiting within this ring or disk of envelope material.  Depending on the detailed core-disk interactions, the core may spiral-in toward the host star and eject the disk material.  This is distinct from the usual CE scenario in that the envelope of the planet becomes bound to the host star, rather than the planet, and it is the planet which ejects its own remnant envelope via tidal and viscous interactions.  Insights into the outcome of such a phase can then be gained from the energy equation
\begin{align}
\alpha_{CE}\left[\frac{GM_*M_{\rm c}}{2a_{\rm f}}-\frac{GM_*M_{\rm c}}{2a_{\rm i}}\right] \simeq  \frac{GM_{*}M_{\rm env}}{2 a_{\rm i}}.
\label{eq:CE}
\end{align}
Here, $\alpha_{\rm CE}$ represents the efficiency with which the planet core's orbital energy can be used to unbind the envelope material which is now in a ring around the host star. The parameters $a_{\rm i}$ and $a_{\rm f}$ denote the orbital separation at the onset and at the end of this `CE' phase, respectively.  Eqn.~(\ref{eq:CE}) can be directly solved for the ratio of $a_f/a_i$, and we find:
\begin{align}
\frac{a_f}{a_i} \simeq \frac{\alpha_{\rm CE} M_{\rm c}}{\alpha_{\rm CE} M_{\rm c}+M_{\rm env}} ~.
\label{eq:CE2}
\end{align}
For plausible values of $\alpha_{\rm CE}$ near unity, this expression yields $a_f/a_i \simeq M_c/M_{\rm pl}$.  This, in turn, implies that the orbital separation would decay by a factor of more than an order of magnitude if the mass transfer is unstable at the onset of RLO.  

Thus, all cores considered in this work would reach their own Roche limit in the event of unstable RLO\footnote{If $R_*$ is too large, the planet-core will plunge into its atmosphere before filling its Roche lobe; the critical period for tidal breakup of a rocky body can be $\lesssim$5 hours~\citep{RappaportSORLW2013}.}.  A new RLO phase would then begin driven, this time, by the interplay of viscous drag on the core due to the remaining envelope mass orbiting the host, and the back-reaction from continuing (now stable) MT through RLO. The former would tend to shrink the orbit, while the latter would cause the orbit to expand.

The actual outcome of an unstable MT phase from a hot Jupiter to its stellar host is still an unexplored question which we are currently addressing via smoothed-particle hydrodynamics simulations. Furthermore, magnetohydrodynamics (MHD) simulations are needed to determine the role played by magnetic fields. Here we note that \cite{Matsakos+15} recently investigated magnetized star-planet interactions via 3D MHD numerical simulations including stellar wind and photo-evaporative mass loss. They found that, depending on the planet's magnetic field and outflow rate, as well as the stellar gravitational field, the star can accrete part of the mass lost by the planet via photo-evaporation. \cite{Pillitteri+15} proved this to be a plausible scenario via far-ultraviolet observations of HD 189733.

\subsection{Planetary Tides}\label{Planetary Tides}
In this work we have taken planetary tides to be efficient in keeping the spin of the planet tidally locked. This assumption is justified by the magnitude of the tidal synchronization timescales discussed in Section~\ref{Orbital Evolution Model}. However, we neglected the resulting effects of tidal dissipation on the planetary structure. These include potentially significant heating, inflation, and resultant stronger mass loss. To gain some intuition into their significance we compute the power deposited between two consecutive integration steps and compare it with the planet luminosity, $L_{\rm pl}$, at each integration step. For the former we use $L_{\rm tide}=\frac{1}{2}I_{\rm pl}|\Omega_{\rm o}^{2}(t+\Delta\,t)-\Omega_{\rm o}^{2}(t)|/\Delta\,t$, where $I_{\rm pl}$ is the moment of inertia.
%, while for the latter we use $E_{\rm bind}=\eta G M_{\rm pl}^{2}/R_{\rm pl}$\footnote{\ron I had asked in my e-mail why this tells us anything since $\Delta\,E_{\rm tide}$ will simply be proportional to $\Delta t$/}. Assuming a uniform density planet ($\eta=3/5$),
We find that $L_{\rm tide}/L_{\rm pl}$ decreases from $10^{-4}$ to $10^{-11}$ throughout the calculation. 
%A core would increase the central density, thus decreasing $\Delta\,E_{\rm tide}$ even further. 
This suggests that the power deposited in the planet via tides may not affect its structure dramatically.

\subsection{Observational Signatures}\label{Obser_Signat}
\begin{figure} [!h]
\epsscale{1.1}
\plotone{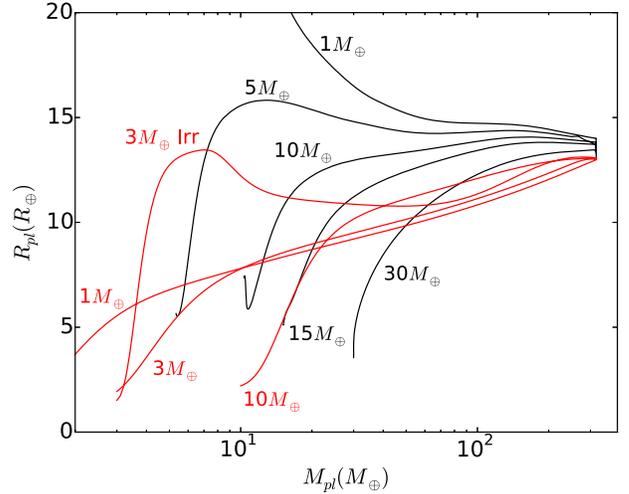}
\caption{Mass-radius relations. In black are the planetary models computed in this work, while in red are those used in \cite{ValsecchiRS2014}.}
\label{fig:Rpl_vs_Mpl_delta0} 
\end{figure}

As discussed in \cite{ValsecchiRS2014}, this evolutionary scenario has several observational consequences. First, the properties of stars hosting hot Jupiters and those hosting Super-Earth and mini-Neptune-type planets should be similar. For example, if all RLO material is lost from the system, the host stars should have similar mass and composition, but super-Earth and mini-Neptune host stars should be somewhat more evolved.
%{\ron (Why??)}.  
Furthermore, depending on the core mass and the details of the MT process (whether any mass is lost from the system and the location where angular momentum is removed), a system may spend enough time in RLO that it might be possible to observe planets in such a phase.
If MT really proceeds through an accretion disk, this may produce observational signatures (e.g., line absorption of stellar radiation and time-dependent obscuration of the starlight; \citealt{LaiHvdH2010}). Finally, the results in Figures~\ref{fig:Mpl_vs_Porb_sigma1gcmM2_delta0_2Gyr_Ftid50epsilon0p1Rxuv0p2percent_separateJdot_ls_MB} and \ref{fig:Mpl_vs_Porb_sigma1gcmM2_multi_2Gyr_Ftid50epsilon0p1Rxuv0p2percent_separateJdot_ls_MB}, and in Table~\ref{Tab:SummaryResults} suggest that, if this model is a viable formation channel for super-Earth and mini-Neptune-type planets, there should be a correlation between $M_{\rm pl}$ and $P_{\rm orb}$. Specifically, the more massive planets should be found at shorter orbital periods. This is a major result when compared to the simple model of~\cite{ValsecchiRS2014}. In fact, without a self-consistent calculation of planetary evolution and irradiation effects for a spectrum of core masses, we had found no trend between final $M_{\rm pl}-P_{\rm orb}$ pairs. 
 The different selection of core masses adopted in \cite{ValsecchiRS2014} does not allow for a one-to-one comparison of those results with the findings of this paper. However these different results can be understood simply via the way in which the orbital separation evolves in response to planetary mass loss, and via the incomplete description of the pre-determined mass-radius relations utilized in ~\cite{ValsecchiRS2014}. The latter fixed $M_{\rm pl}-R_{\rm pl}$ relations are shown in Figure~\ref{fig:Rpl_vs_Mpl_delta0}, compared with those {\it computed} with MESA during the course of the evolutions.
%%%%%%%%%%%%%%%%%%%%%%
%%%%%%%%%%%%%%%%%%%%%%

This inverse correlation between orbital period and planetary mass that we have found also suggests that dynamically stable MT phases like those presented here do not seem to represent a viable channel for the formation of the so-called `ultra-short-period planets' (e.g., `USPs'; \citealt{SanchisOjedaRWKLE2014} and reference therein). These planets have typical radii smaller than 2$R_{\oplus}$ (corresponding to a mass of about 5\,$M_{\oplus}$; \citealt{WeissMarcy2014}) and orbital periods shorter than one day.
%, Earth and super-Earth type planets ($R_{\rm pl}\leq\,R_{\oplus}$) in orbital periods less than 1\, d 
\\
\\
%%%%%%%%%%%%%%%%%%%%%%
\section{Conclusions} \label{Conclusions}
In this paper we have presented the first evolutionary calculations of irradiated hot-Jupiters undergoing tidally-driven Roche-lobe overflow (RLO).  We found that, depending on the size of the planetary core and the details of the mass transfer, the RLO phase and, in turn, the detached phase (after RLO had ceased) can last from a few to several Gyrs. For the smaller core masses ($M_{\rm c}\leq\,15\,M_{\oplus}$), after most of the envelope has been removed during RLO, the planet spends a few Gyrs losing mass via photo-evaporation at nearly constant orbital period. This is consistent with the density of known super-Earths and sub-Neptunes, for which detailed modeling \citep{LopezFortney2014} place tens of percent of the total mass in a H/He envelope surrounding a rocky core.

As noted above, we find an inverse correlation between the core mass of the planet and its final orbital period.  This results from the basic fact that, in general, the irradiated planets with larger core masses decrease in radius faster/earlier with mass loss than planets with lower-mass cores.  Final orbital periods of $\lesssim$ 1 day appear to require large core masses of $\gtrsim 15 \,M_\oplus$, which are likely substantially higher than the masses of the USP planets found by \citet{SanchisOjedaRWKLE2014}. Thus, the scenario we are presenting here probably does not account for the USPs.

In this work we have considered a coarse grid of core masses and {\it one} initial binary configuration. However, as summarized in Tables~1 and 2 of \citealt{ValsecchiR2014b},  the tightest observed hot-Jupiter systems comprise a variety of stellar and planetary masses ($\simeq\,0.87-1.33\,M_{\odot}$ and $\simeq\,0.46-1.49\,M_{\rm J}$), as well as metallicities ($Fe/H\simeq-0.35-0.22$) and ages (1.5$-$13\,Gyr). All these parameters may affect the efficiency of tides and, thus, the evolution of a Jupiter undergoing RLO. This variety of properties requires exploration of a more refined grid in parameter space of initial component and orbital properties, as well as planetary core masses (especially in the $M_{\rm c}\,\leq\,15\,M_{\oplus}$ regime; Figures~\ref{fig:Mpl_vs_Porb_sigma1gcmM2_delta0_2Gyr_Ftid50epsilon0p1Rxuv0p2percent_separateJdot_ls_MB} and \ref{fig:Mpl_vs_Porb_sigma1gcmM2_multi_2Gyr_Ftid50epsilon0p1Rxuv0p2percent_separateJdot_ls_MB}). Such an extended parameter space study should also explore in more detail the boundaries for MT stability.  
Future observations of increasingly massive super-Earths and sub-Neptune-type planets in increasingly tighter orbits might provide an important observational test of the ideas presented here.

Finally, we remark that the value of $\gamma \delta$, the parameter describing mass and specific angular momentum loss, is crucial to the stability of RLO mass transfer.  Where matter goes after RLO, and how much of it is actually accreted by the host star or ejected from the system, can only be determined by hydrodynamic calculations.  The results of such calculations could well determine whether hot Jupiters can undergo the kind of stable RLO mass transfer described in this work.

%%%%%%%%%%%
------------------------------------------
 \acknowledgments

\begin{acknowledgements}
We thank the anonymous referee for many very helpful comments. FV and FAR are supported by NASA Grant NNX12AI86G. FV is also supported by a CIERA fellowship. LAR gratefully acknowledges support provided by NASA through Hubble Fellowship grant \#HF-51313 awarded by the Space Telescope Science Institute, which is operated by the Association of Universities for Research in Astronomy, Inc., for NASA, under contract NAS 5-26555.  We thank Dorian Abbot, Arieh Konigl, Titos Matsakos, Ruth Murray-Clay, James Owen, and Dave Stevenson for useful discussions. This work used computing resources at CIERA funded by NSF PHY-1126812. This research has made use of the NASA Exoplanet Archive, which is operated by the California Institute of Technology, under contract with the National Aeronautics and Space Administration under the Exoplanet Exploration Program.
\end{acknowledgements}

\bibliography{myBibtex}{}
\bibliographystyle{apj}

%%%%%%%%%%%%%%%%%%%%%%%%%%
\newpage
\clearpage
\begin{sidewaystable}
\centering
\caption{Summary of Results.}
\begin{tabular}{c|c|c|c|c|c|c|c|c|c|c|c|ccc}
\hline 
$M_{\rm c}$ & $\gamma$&$\Sigma_{\rm pl}$ & PE & $\Delta$t$_{\rm RLO}$ & $P_{\rm orb}$ at the & $f$ at the &$\Delta$t when & $P_{\rm orb}$ when&$\Delta$t when & $P_{\rm orb}$ when & $t$ when\\
&& & &  & end of RLO & end of RLO & $f\sim$ 20\% to 30\% & $f\sim$20\%& $f\sim$7\% to 20\% & $f\sim$7\% & $f\sim$7\%&\\
($M_{\oplus}$) & &  (g\,cm$^{-2}$) & &(Gyr) & (d) &  (\%) & (Gyr) & (d) & (Gyr) & (d)& ($t_{\rm MS}$)\\													 
\\[-1.0em]
\hline
5  & $-$ & 1 & e-lim & 4.5 & 3.4 & 39.8 & 0.2 & 3.4 & 3.1 & 3.4 & 1.0\\ 
10& $-$ & 1 & e-lim& 4.4 & 1.6 & 43.1 & 0.5 & 1.6 & 2.9 & 1.6 & 1.0\\
15& $-$ & 1 & e-lim& 4.2 & 1.2 & 43.1 & 0.6 & 1.1 & 2.0 & 0.9 &0.9 \\
30& $-$ & 1 & e-lim& 3.7 & 0.5 & 7.0 & 0.3 & 0.6 & 0.3 & 0.5 & 0.6\\
%{\bf 30}& $-$ & {\bf 1} & {\bf 5.7} & {\bf 0.3} & {\bf 1.0} & {\bf 0.3} & {\bf 0.5} & {\bf 0.1} & {\bf 0.4} & {\bf 0.8}&$-$\\
\hline
5& 0.6 & 1 &e-lim & 1.6 & 3.9 & 41.1 & 0.1 & 3.9 & 5.2 & 3.9 & 0.9\\
10& 0.7 & 1 & e-lim& 1.2 & 1.7 & 46.0 & 0.3 & 1.7 & 3.4 & 1.7 & 0.7\\
15 & 0.7 & 1 &e-lim & 1.3 & 1.2 & 48.2 & 0.5 & 1.2 & 3.4 & 1.1 & 0.8\\
30 & 0.7 & 1 &e-lim & 2.0 & 0.5 & 7.0 & 0.3 & 0.6 & 0.4 & 0.5 & 0.4\\
\hline
5  & $-$ & 1 & e-lim + rr-lim & 6.9 & 3.8 & 37.7 & 0.2 & 3.8 & 1.6 & 3.8 & 1.1\\ 
30& $-$ & 1 & e-lim + rr-lim & 5.6 & 0.5 & 7.0 & 0.4 & 0.6 & 0.3 & 0.5 & 0.8\\
5& 0.6 & 1 &e-lim + rr-lim & 3.7 & 3.5 & 40.1 & 0.2 & 3.5 & 3.8 & 3.5 & 1.0 \\
30 & 0.7 & 1 &e-lim + rr-lim & 3.0 & 0.5 & 7.0 & 0.4 & 0.6 & 0.3 & 0.5 & 0.5\\
%{\bf 30}& {\bf 0.7} & {\bf 1} & {\bf 3.1} & {\bf 0.3} & {\bf 1.0} & {\bf 0.3} & {\bf 0.5} & {\bf 0.1} & {\bf 0.4} & {\bf 0.5}&$-$\\
\hline
5  & $-$ & 100 & e-lim & 4.7 & 3.7 & 40.2 & 0.2 & 3.7 & 3.5 & 3.7 & 1.0\\ 
30& $-$ & 100 & e-lim& 4.2 & 0.5 & 7.0 & 0.4 & 0.7 & 0.4 & 0.5 & 0.6\\
5& 0.6 & 100 &e-lim & 1.7 & 4.3 & 41.7 & 0.2 & 4.3 & {\bf 5.3} & {\bf 4.3} & {\bf 1.0} \\
30 & 0.7 & 100 &e-lim & 2.0 & 0.7 & 19.4 & 0.3 & 0.7 & 0.6 & 0.5 & 0.5 \\
%below are the results with minimum mass fraction considered in the envelope set to 1 percent or the minimum we can run the model to
%5  & 0 & 1 & 6.9 & 3.8 & 38.1 & 1.3 & 3.8 & 0.5 & 3.6 & 1.1 &$-$\\
%5\footnote{The minimum $f$ value considered is $\simeq$3.9\%.}  & 0 & 100 & 7.2 & 4.2 & 37.7 & 1.2 & 4.2 & 0.1 & 4.2 & 1.1 &$-$\\
%10& 0 & 1 & 6.9 & 1.7 & 39.7 & 1.0 & 1.6 & 0.2 & 0.9 & 1.1&10.9\\
%15& 0 & 1 & 6.6 & 1.2 & 36.4 & 0.6 & 0.9 & 0.3 & 0.5 &1.0 &10.3\\
%30& 0 & 1 & 5.7 & 0.3 & 1 & 0.3 & 0.5 & 0.1 & 0.3 & 0.8&$-$\\
%30& 0 & 100 & 6.2 & 0.4 & 1 & 0.3 & 0.6 & 0.2 & 0.4 & 0.8&$-$\\
%5\footnote{The minimum $f$ value considered is $\simeq$2\%.}  & 0.6 & 1 & 3.6 & 3.5 & 40.1 & 2.0 & 3.5 & 3.0 & 3.4 & 1.1 &$-$\\
%10\footnote{The minimum $f$ value considered is $\simeq$2.6\%.}  & 0.7 & 1 & 3.3 & 1.6 & 43.7 & 1.8 & 1.6 & 3.0 & 1.0 & 1.1 &10.9\\
%15 & 0.7 & 1 & 3.6 & 1.2 & 43.7 & 1.5 & 1.0 & 1.6 & 0.5 & 1.0 &$-$\\
%30 & 0.7 & 1 & 3.1 & 0.3 & 1.0 & 0.3 & 0.5 & 0.1 & 0.3 & 0.5 &$-$\\
\\[-1.0em]
\hline \\[-1.0em]
\end{tabular}
\label{Tab:SummaryResults}
\tablecomments{The results include $f\,=\,M_{\rm env}/M_{\rm pl}$ values down to 7\% (but for the values in boldface where $f\geq$8\%), to allow a comparison between different MT assumptions and $M_{\rm c}$ values. In fact, for some of the models considered MESA does not converge when $M_{\rm env}/M_{\rm pl}$ drops below about 7\%. Time intervals are denoted with $\Delta$\,t. Column $PE$ denotes the photo-evaporation prescription, with $e-lim$ and $rr-lim$ indicating the energy-limited and radiation/recombination-limited regimes, respectively. The parameter $t$ denote the age of the system in units of the stellar main sequence lifetime. The first four examples are for conservative MT. For non-conservative MT we use $\delta\,=\,1$ and vary $\gamma$ as summarized in the second column. Note, the 30\,$M_{\oplus}$ core cases never detaches, but for the last case (non-conservative MT evolution with $\Sigma_{\rm pl} 100\,$g~cm$^{-2}$). In this case, 0.5\,Gyr after the first 2\,Gyr long RLO phase, the planet  fills its Roche lobe again). 
%{\bon In boldface are the results including stellar wind mass loss for the 30\,$M_{\oplus}$ core case, to show that there is no significant difference}.
}
\end{sidewaystable}

%%%%%%%%%%%%%%%%%%%%%%%%%%

\end{document}